\documentclass[twocolumn,trackchanges]{aastex701}

\usepackage{amsmath}
\usepackage{placeins} 
\usepackage{afterpage} 

\definecolor{blue1}{RGB}{51, 153, 255} 
\definecolor{pink1}{RGB}{255, 102, 178}
\definecolor{pink2}{RGB}{255, 153, 255} 
\definecolor{purple1}{RGB}{204, 153, 255} 
\definecolor{purple2}{RGB}{153, 51, 255} 
\definecolor{blue1}{RGB}{51, 153, 255} 

\newcommand{\reff}{$R_{\text{eff}}$}
\newcommand{\Ha}{H$\alpha$}

\newcommand{\OIIIHb}{[OIII] + H$\beta$}

\newcommand{\pz}{photo-$z$}
\newcommand{\snr}{$S/N$}

\usepackage{float} 
\begin{document}

\title{A Population of Red Galaxies with Very Strong Emission Lines at $z > 5$ Revealed by the NIRCam Medium Bands: ``Classic'' LRDs, Dusty Star-Forming Galaxies, and a Missing Population of LRDs}

\author[0009-0000-8716-7695]{Sunna Withers}
\affiliation{Department of Physics and Astronomy, York University, 4700 Keele St. Toronto, Ontario, M3J 1P3, Canada\\}
\email{sunnaw@my.yorku.ca}

\author[0000-0002-9330-9108]{Adam Muzzin} 
\affiliation{Department of Physics and Astronomy, York University, 4700 Keele St. Toronto, Ontario, M3J 1P3, Canada\\}
\email{muzzin@yorku.ca}

\author[0000-0002-5269-6527]{Swara Ravindranath}
\affiliation{Astrophysics Science Division, NASA Goddard Space Flight Center, 8800 Greenbelt Road, Greenbelt, MD 20771, USA\\}
\affiliation{Center for Research and Exploration in Space Science \& Technology II, Department of Physics, The Catholic University of
America, 620 Michigan Ave., N.E. Washington, DC 20064, USA\\}
\email{swara.ravindranath@nasa.gov}

\author[0000-0002-4201-7367]{Chris J. Willott}
\affiliation{National Research Council of Canada, Herzberg Astronomy \& Astrophysics Research Centre, 5071 West Saanich Road, Victoria, BC, V9E 2E7, Canada\\}
\email{chriswillott1@gmail.com}

\author[0000-0003-3243-9969]{Nicholas S. Martis}
\affiliation{Faculty of Mathematics and Physics, Jadranska ulica 19, SI-1000 Ljubljana, Slovenia\\}
\email{nicholas.martis@fmf.uni-lj.si}

\author[0000-0002-9909-3491]{Roberta Tripodi}
\affiliation{INAF - Osservatorio Astronomico di Roma, Via Frascati 33, Monte Porzio Catone, 00078, Italy\\}
\email{roberta.tripodi@inaf.it}

\author[0000-0003-3983-5438]{Yoshihisa Asada}
\affiliation{Dunlap Institute for Astronomy and Astrophysics, 50 St. George Street, Toronto, Ontario, M5S 3H4, Canada\\}
\email{yoshi.asada@utoronto.ca}

\author[0000-0001-5984-0395]{Maru\v{s}a Brada\v{c}}
\affiliation{Faculty of Mathematics and Physics, Jadranska ulica 19, SI-1000 Ljubljana, Slovenia\\}
\affiliation{Department of Physics and Astronomy, University of California Davis, 1 Shields Avenue, Davis, CA 95616, USA\\}
\email{marusa.bradac@fmf.uni-lj.si}

\author[0009-0000-5385-8674]{Maya Merchant}
\affiliation{Department of Physics and Astronomy, York University, 4700 Keele St. Toronto, Ontario, M3J 1P3, Canada\\}
\affiliation{Niels Bohr Institute, University of Copenhagen, Jagtvej 128, DK-2200 Copenhagen N, Denmark\\}
\email{merchm@yorku.ca}

\author[0000-0002-8530-9765]{Lamiya Mowla}
\affiliation{Whitin Observatory, Department of Physics and Astronomy, Wellesley College, 106 Central Street, Wellesley, MA 02481, USA\\}
\email{lmowla@wellesley.edu}

\author[]{Ga\"el Noirot}
\affiliation{Space Telescope Science Institute, 3700 San Martin Drive, Baltimore, Maryland 21218, USA\\}
\email{gnoirot@stsci.edu}

\author[0000-0001-8830-2166]{Ghassan T. E. Sarrouh}
\affiliation{Department of Physics and Astronomy, York University, 4700 Keele St. Toronto, Ontario, M3J 1P3, Canada\\}
\email{gsarrouh@yorku.ca}

\author[0000-0002-7712-7857]{Marcin Sawicki}
\affiliation{Department of Astronomy and Physics and Institute for Computational Astrophysics, Saint Mary's University, 923 Robie Street, Halifax, Nova Scotia B3H 3C3, Canada}
\email{marcin.sawicki@smu.ca}

\author[0000-0002-0243-6575]{Jacqueline Antwi-Danso}
\affiliation{David A. Dunlap Department of Astronomy and Astrophysics, University of Toronto, 50 St. George Street, Toronto, Ontario, M5S 3H4, Canada\\}
\email{j.antwidanso@utoronto.ca}

\author[0000-0001-9414-6382]{Anishya Harshan}
\affiliation{Kavli Institute for Cosmology, University of Cambridge, Madingley Road, Cambridge, CB3 0HA, United Kingdom\\}
\affiliation{Cavendish Laboratory - Astrophysics Group, University of Cambridge, 19 JJ Thomson Avenue, Cambridge, CB3 0HE, United Kingdom\\}
\email{at2207@cam.ac.uk}

\author[0009-0009-9848-3074]{Naadiyah Jagga}
\affiliation{Department of Physics and Astronomy, York University, 4700 Keele St. Toronto, Ontario, M3J 1P3, Canada\\}
\email{jagga@yorku.ca}

\author[0000-0001-9002-3502]{Danilo Marchesini}
\affiliation{Department of Physics and Astronomy, Tufts University, 574 Boston Avenue, Suite 304, Medford, MA 02155, USA\\}
\email{danilo.marchesini@tufts.edu}

\author[0009-0009-2307-2350]{Katherine Myers}
\affiliation{Department of Physics and Astronomy, York University, 4700 Keele St. Toronto, Ontario, M3J 1P3, Canada\\}
\email{kjmyers@yorku.ca}

\author[0000-0003-0780-9526]{Visal Sok}
\affiliation{Department of Astrophysical and Planetary Sciences, University of Colorado, 2000 Colorado Ave, Boulder, CO 80309, USA\\}
\email{visal.sok@colorado.edu}



\begin{abstract}

The NIRCam medium-bands have proven to be efficient at identifying Emission Line Galaxies (ELGs) with high equivalent width (EW) \Ha\ and \OIIIHb\ emission lines. In this paper we exploit this efficiency to identify a sample of ELGs at $4.9 \lesssim z \lesssim 8.9$ using medium-band imaging from the CANUCS, Technicolor, and JUMPS surveys. We find that the ELGs exhibit a strong correlation between continuum color and emission line strength, such that galaxies with bluer UV/optical continua have stronger \Ha\ and \OIIIHb\ emission lines. We identify 26 galaxies that are outliers from this relation, which we call the Red Emission line Galaxies (REGs), because of their red continuum color and strong emission lines. We classify the REGs into three categories: 1) “classic” Little Red Dots (LRDs) selected with common literature criteria, 2) extended REGs, resolved in $F444W$ and consistent with being Dusty Star Forming Galaxies (DSFGs), and 3) compact REGs, unresolved in $F444W$ but not classified as LRDs. The compact REGs fail common LRD selections for several reasons, including faint continuua, contamination from very strong \OIIIHb, and UV/optical colors that are flatter than those of LRDs. We conclude that the compact REGs are likely LRDs that “classic” selection criteria miss, and are therefore missing from existing samples. Our results suggest that medium-band selection can provide more complete samples of these objects.

\end{abstract}



\section{Introduction} \label{sec:intro}

The James Webb Space Telescope (JWST) has provided extensive observations galaxies at $ z > 5$, revealing the high-redshift galaxy population to be remarkably diverse. One of the most abundant populations of high-redshift galaxies are emission-line galaxies (ELGs, e.g. \citealt{williams_jems_2023}, \citealt{matthee_eiger_2023} \citealt{boyett_extreme_2024}, \citealt{guo_search_2025}, \citealt{llerena_extreme_2025}), which have high equivalent-width (EW) optical emission lines driven by star formation or active galactic nuclei (AGN). While there is no universal definition of ELGs, the \Ha\ and \OIIIHb\ EWs of ELGs at $z \gtrsim 5$ commonly exceed $\gtrsim 500$\AA, even reaching $\gtrsim 2000$\AA\ in the most extreme cases (e.g. \citealt{withers_spectroscopy_2023}, \citealt{davis_census_2023}, \citealt{trussler_cloudy_2025}). 

Beyond ELGs, which often exhibit blue continuum colors (e.g. \citealt{harshan_canucs_2024}, \citealt{felicioni_uv_2026}), JWST has identified a substantial population of very red galaxies at $z > 3$. This population includes extremely red objects (EROs), and consists of several types of galaxies, including quiescent galaxies, dusty star forming galaxies (DSFGs), and AGN/Little Red Dots (LRDs, e.g. \citealt{nelson_jwst_2023}, \citealt{rodighiero_jwst_2023}, \citealt{gottumukkala_unveiling_2024}, \citealt{williams_galaxies_2024}, \citealt{barro_extremely_2024}, \citealt{antwi-danso_feniks_2025}, \citealt{kocevski_rise_2025}, \citealt{tripodi_extreme_2025}). As such, red galaxies have a wide range of \Ha\ and \OIIIHb\ EWs. While many have weak to moderate EW emission lines (e.g. \citealt{barrufet_strength_2025} find EW(\Ha) $\sim 68 - 550 $\AA), a subset have much higher EWs and qualify as ELGs (e.g. \citealt{perez-gonzalez_ceers_2023}, \citealt{akins_two_2023}).

Galaxies with both high EW emission lines and red continuum colors represent a unique and extreme subset of the overall galaxy population. Many of these red ELGs have the ``V-shaped'' spectral energy distributions (SEDs) characteristic of LRDs (e.g. XELG-$z6$ of \citealt{perez-gonzalez_ceers_2023}, \citealt{kocevski_rise_2025}, \citealt{setton_little_2025}), while others lack this UV excess and are red due to large amounts of dust attenuation, with $A_V > 1$ mag (e.g. \citealt{martis_canucstechnicolor_2025}). The current interpretation of these red ELGs is that they host obscured starbursts or AGN, and constitute some of the most massive and vigorously star-forming systems at high-redshift. Additionally, there is evidence that high-redshift dust obscured galaxies are progenitors to later populations of massive or extreme galaxies (e.g. \citealt{akins_two_2023}, \citealt{tarrasse_compact_2025}, \citealt{ganguly_stellar_2025}). As such, red galaxies with strong emission lines are a valuable tool for probing early galaxy evolution, as well as obscured star formation and AGN at high-redshift. 

Many existing samples of red galaxies at $z > 3$ have been identified through broad-band color selections, such as $F150W - F444W$ or $F150W - F356W$ color cuts (e.g. \citealt{barro_extremely_2024}, \citealt{barrufet_quiescent_2025}, \citealt{yang_unveiling_2025}). While these selections have produced large samples of red galaxies, they often place signal-to-noise ratio (\snr) or magnitude cuts on at least one broad-band filter (often $F444W_{mag} \lesssim 27 - 28$ mag, e.g. \citealt{barrufet_strength_2025}), which biases samples towards brighter galaxies. The population of faint, red galaxies could represent an important sub-population of red galaxies, but such objects remain poorly understood due to limited sample sizes (e.g. \citealt{akins_two_2023}, \citealt{rodighiero_jwst_2023}, \citealt{williams_galaxies_2024}, \citealt{gottumukkala_unveiling_2024}).

An effective way to probe populations of faint red galaxies is through the NIRCam medium-band filters. Not only does the increased spectral resolution of the medium-bands provide more accurate estimates of galaxy redshifts, physical, and spectral properties (e.g. \citealt{asada_bursty_2023}, \citealt{sarrouh_exposing_2024}, \citealt{lorenz_measuring_2025}), they also enable robust selection of galaxies based on their emission lines. Such selections often involve medium-band color cuts that target flux excesses driven by strong emission lines, most commonly the \Ha\ or \OIIIHb\ lines. Existing medium-band selected samples of ELGs have already proven useful at investigating red, high-redshift galaxies. For example, \citealt{martis_canucstechnicolor_2025} used a sample of medium-band selected ELGs to investigate obscured star-formation at $z > 7.5$, and highlighted the effectiveness of such samples at probing obscured systems.

In this paper, we use NIRCam medium-band imaging to identify 26 galaxies with red continuum fluxes and strong \Ha\ or \OIIIHb\ emission lines at $z \gtrsim 5$, which we call the Red Emission line Galaxies (REGs). We classify the REGs into three categories: (1) classic LRDs, (2) extended REGs, and (3) compact REGs. Our interpretations of the classic LRDs and the extended REGs are straightforward, however we show the interpretation of the compact REGs is more puzzling. Overall, we conclude the compact REGs likely constitute a population of LRDs which fail typical selection criteria (e.g. \citealt{kocevski_rise_2025}, \citealt{kokorev_census_2024}) and are missing from current samples. We discuss our observations in \S \ref{sec:observations}, our sample selections in \S \ref{sec:reference_selection} and \S \ref{sec:reg_selection}, and classification in \S \ref{sec:Nature_REGs}. Finally, we discuss the nature of the compact REGs in more detail in \S \ref{sec:discussion}. We assume a cosmology of $H_0 = 72 \text{ km s}^{-1} \text{Mpc}^{-1}$, $\Omega_m = 0.27$ and $\Omega_\Lambda = 0.73$. All magnitudes are reported using the AB magnitude system (\citealt{oke_absolute_1974}, \citealt{oke_secondary_1983}), and all EWs are reported in the rest-frame. 

\begin{figure*}
    \centering
    \includegraphics[width=0.9   \linewidth]{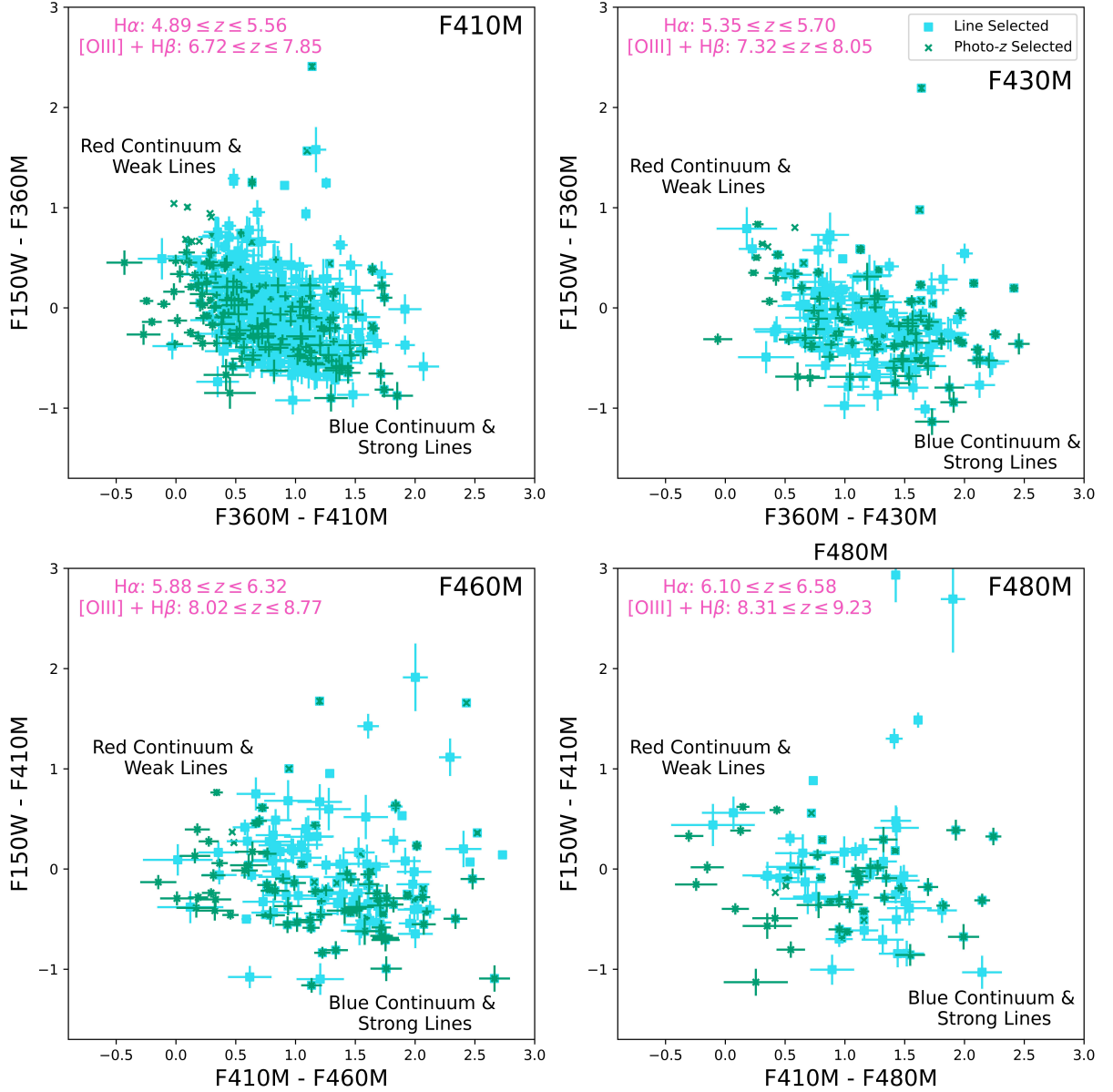}
    \caption{Color-color diagrams for galaxies with the \Ha\ or \OIIIHb\ emission lines in one of the $F410M$ (top left), $F430M$ (top right), $F460M$ (bottom left), and $F480M$ (bottom right) bands, with the sample filtered to galaxies at the correct redshift (see also Table \ref{tab:Selection_Summary}). In each panel, the $y-$axis color highlights the rest-frame UV-optical continuum, using $F150W$ to measure the rest-frame UV, and either $F360M$ or $F410M$ to measure the rest-frame optical (see Table \ref{tab:Selection_Summary}). The $x$-axis color then highlights the strength of the \Ha\ or \OIIIHb\ emission lines, using either $F360M$ or $F410M$ to measure the rest-frame optical continuum, and the band containing an emission line. Bluer $x-$axis colors indicate stronger emission lines while redder colors indicate weaker lines.}
    \label{fig:ColourColour_byz}
\end{figure*}

\section{Observations} \label{sec:observations}

We use NIRCam imaging obtained as part of the CAnadian NIRISS Unbiased Cluster Survey (CANUCS, \citealt{willott_near-infrared_2022}), JWST in Technicolor (\citealt{sarrouh_canucstechnicolor_2026}), and JWST Ultimate Medium-band Photometric Survey (JUMPS, \citealt{withers_jumps_2024}, GO ID 5890) programs. CANUCS provides deep NIRCam imaging ($\sim 28.9$ mag in medium bands, $\sim 29.4$ mag in broad bands to $5 \sigma$ for point sources) of ten fields: five massive galaxy cluster fields (Abell 370, MACS0416, MACS0417, MACS1149, and MACS1423), and five adjacent flanking fields. The cluster fields have observations in seven broad and one medium-band filters ($F090W$, $F115W$, $F150W$, $F200W$, $F277W$, $F356W$, $F410M$, and $F444W$), while the flanking fields have imaging in five broad-band ($F090W$, $F115W$, $F150W$, $F277W$, $F444W$) and nine medium-band ($F140M$, $F162M$, $F182M$, $F210M$, $F250M$, $F300M$, $F335M$, $F360M$, $F410M$) filters. JWST in Technicolor then provides additional imaging in three of the flanking fields (adjacent to Abell 370, MACS0416, MACS1149) in three broad-band ($F070W$, $F200W$, $F356W$), three medium-band ($F430M$, $F460M$, $F480M$), and two narrow-band ($F164N$, $F187N$) filters. Additionally, JUMPS provides imaging of three cluster fields (Abell 370, MACS0416, MACS1149) in four medium-bands ($F360M$, $F430M$, $F460M$, and $F480M$) and ultra-deep ($\sim30$ mag to $5\sigma$ for point sources) $F150W$ imaging. The $5\sigma$ depths of the Technicolor and JUMPS medium-band imaging are similar to those of the CANUCS medium-bands

Each of the CANUCS, Technicolor, and JUMPS fields were observed in a single NIRCam pointing using modules A and B. The position angle of the Technicolor and JUMPS observations is matched to that of CANUCS to ensure observations fully overlap. In this work, we utilize all five of the CANUCS flanking fields as they all include extensive medium-band observations, but only include the three CANUCS cluster fields which have JUMPS imaging. This corresponds to $77.6$ arcmin$^2$ of CANUCS observations, including $58.2$ arcmin$^2$ which have additional long-wavelength medium-band imaging provided by Technicolor and JUMPS. The CANUCS, Technicolor, and JUMPS fields also have deep imaging available from the Hubble Space Telescope (HST). The HST imaging varies across each field, including observations from ACS, WFC3/UVIS, and WFC3/IR, provided by \citealt{lotz_frontier_2017}, \citealt{postman_cluster_2012} HST-GO-16667 (PI: M. Bradac), and PID-11507 (PI: K.Noll). Imaging from ACS and WFC3/UVIS are included in our analysis, however, we exclude observations from WFC3/IR owing to their shallower depths relative to available JWST imaging. 

Additionally, CANUCS includes follow-up spectroscopy with the NIRSpec prism, which provides low-resolution ($R\sim 100$) spectroscopy over $0.6\mu\text{m} < \lambda < 5.3\mu\text{m}$. We make use of the spectroscopic observations for the REGs when available. The data processing of NIRCam imaging and NIRSpec prism spectroscopy is discussed in detail in \cite{sarrouh_canucstechnicolor_2026}, including a summary of the HST, CANUCS, and Technicolor data available in each field (Table 1 of \citealt{sarrouh_canucstechnicolor_2026}). The JUMPS observations are not discussed in \cite{sarrouh_canucstechnicolor_2026}, however they were processed using the same techniques as the CANUCS and Technicolor data. 

\section{Reference Sample Selection} \label{sec:reference_selection}

The REGs are a subset of two broader galaxy samples: (1) a sample of ELGs selected using NIRCam medium-band color cuts, the ``line-selected sample'', and (2) a sample of photometric redshift-selected galaxies, the ``\pz-selected sample''. Both samples are similar to those used in \cite{martis_canucstechnicolor_2025}, but extend down to lower redshifts. In this section, we discuss both the line-selected (\S\ref{subsec:line_sample}) and \pz\ selected samples (\S\ref{subsec:pz-sample}), as well as measurements of their physical and spectral properties (\S\ref{subsec:measuring_properties}). Additionally, we compare the sample of REGs to a sample of LRDs in CANUCS, Technicolor, and JUMPS, which we discuss in \S\ref{subsec:LRD_Selections}. 

\subsection{Line Selected Sample} \label{subsec:line_sample}

The line-selected sample consists of ELGs identified from NIRCam medium-band color selections, similar to the sample originally defined in \citealt{withers_spectroscopy_2023}. Full details of the selection will be presented in a forthcoming paper. Here we briefly describe the colors cuts, which were defined following the method of \cite{withers_spectroscopy_2023}. We generate synthetic NIRCam observations of galaxies using publicly available \texttt{Yggrasil} spectral energy distributions (SEDs, \citealt{zackrisson_spectral_2011}) and identify which medium-band colors are dominated by the \Ha\ or \OIIIHb\ emission line complexes as a function of redshift. We then use these to define single-line color cuts which identify galaxies with \Ha\ and \OIIIHb\ emission as a function of redshift, which are sensitive to galaxies with rest-frame EWs $\gtrsim 200$\AA.

By selecting galaxies based on medium-band flux excesses driven by emission lines, we are able to identify galaxies independent of their underlying continuum emission. As such, we can robustly identify populations of galaxies which have low \snr\ in the NIRCam broad-bands, and thus poorly constrained Lyman breaks. In order to ensure such galaxies are included in the line-selected sample, we do not impose any \snr\ cuts on the broad-bands, as is generally required in broad-band selections of high-redshift galaxies (e.g. $F444W$ in ERO selections, \cite{williams_galaxies_2024}). Instead, we require sources have $S/N \geq 3$ in the medium-band containing an emission line, and that medium-band colors have $S/N \geq 4$.

In addition to the color and \snr\ cuts, we utilize several other quality cuts to minimize contamination in our sample. First, we employ a redshift cut using photometric redshifts measured from the \texttt{EAZY-py} SED fitting code (\citealt{brammer_eazy_2008}). The photometric redshifts are measured using the \cite{larson_spectral_2023} templates and \cite{asada_improving_2025} prescription for IGM/CGM attenuation. This redshift cut determines which of the \Ha\ or \OIIIHb\ emission lines is responsible for driving the medium-band colors, and excludes low-redshift interlopers from our sample (e.g. $z \sim 1 - 2$ Pa$-\alpha$ or Pa$-\beta$ emitters). While our sample of ELGs extends down to $z\sim 3$, we limit this analysis to galaxies with \Ha\ or \OIIIHb\ in one of the $F410M$, $F430M$, $F460M$, or $F480M$ bands. This corresponds to redshifts $4.9 \lesssim z \lesssim 9.2$. Additionally, we require galaxies to have observations in HST bands $F435W$ and $F606W$. This further excludes low-redshift interlopers by ensuring our sources are not detected bluewards of the Lyman break. 

Finally, we account for contamination from brown dwarfs using the Sonora brown dwarf templates (\citealt{marley_sonora_2021}). The Sonora templates provide SEDs for brown dwarfs with effective temperatures $200 \text{K}\leq T_{\text{eff}} \leq 2400\text{K}$, which extends down to the extremely cold brown dwarfs shown to contaminate samples of high-redshift galaxies (e.g. \citealt{gandolfi_mysteries_2026}, \citealt{bradac_two_2026}). We fit our sample with the Sonora templates in \texttt{EAZY-py}, and require sources be better fit by galaxy templates than brown dwarf templates by enforcing $\chi^2_{Galaxy} < \chi^2_{BrownDwarf}$. 

We visually inspect the sample to exclude spurious sources, such as cosmic rays near the detector edges. Following this visual inspection, these cuts produce a sample of of 2060 ELGs over $4.89 < z < 9.28$. There are 86 objects in the line-selected sample which have available spectroscopic redshifts from CANUCS NIRSpec prism observations. These yield a $|z_{spec} - z_{phot}| = 0.015$, confirming the accuracy of the medium-band color selections. 

\begin{figure*}
    \centering
    \includegraphics[width=0.85\linewidth]{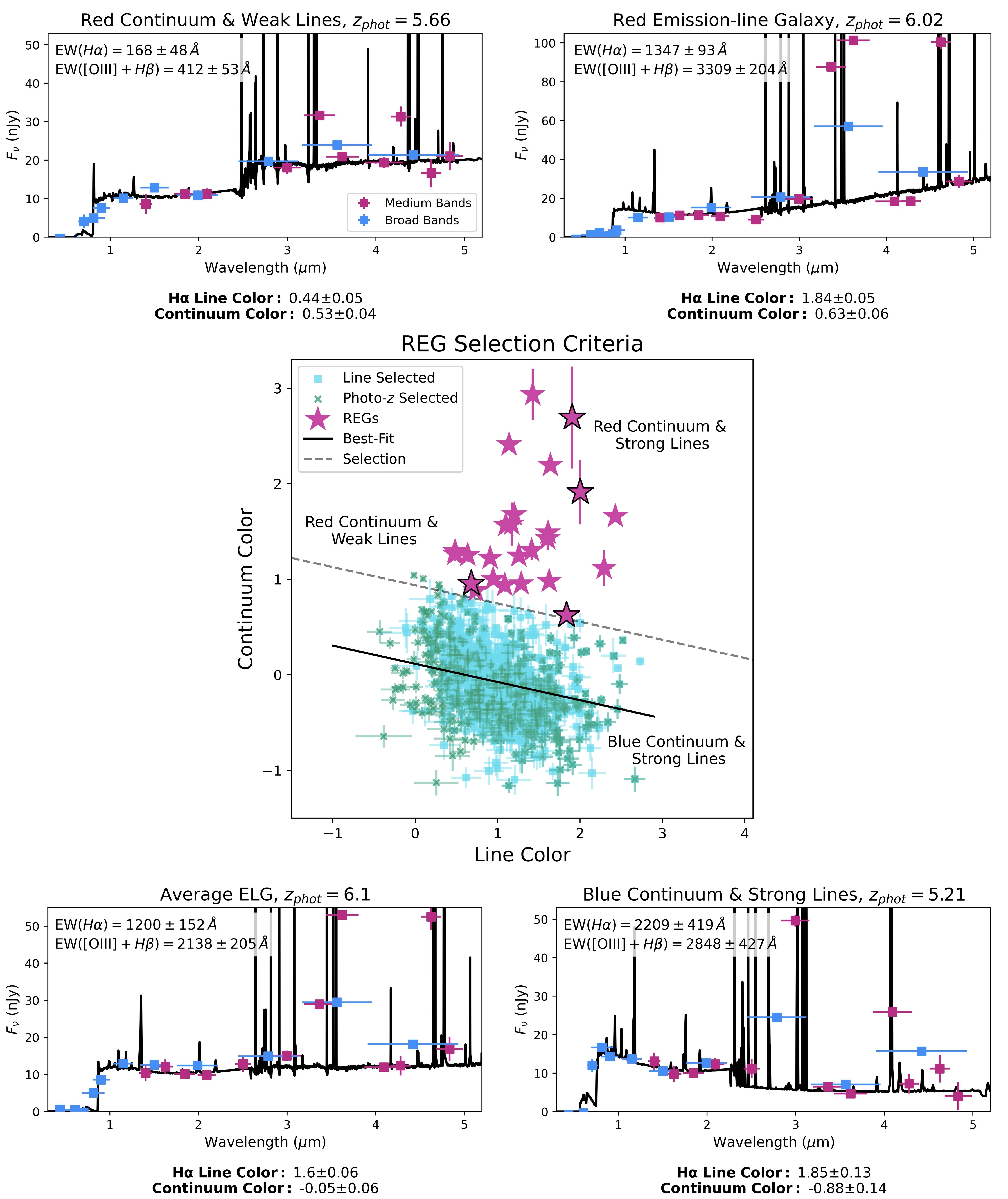}
    \caption{All four panels in Figure \ref{fig:ColourColour_byz} combined onto one panel. There is a correlation between emission line strength and continuum color, where bluer galaxies have stronger emission lines (bottom right), and redder galaxies have weaker emission lines (top left). We fit both samples with a straight line which allows us to calculate an expected continuum color based on a given emission line color. We then define the REGs (pink stars) as galaxies with continuum colors $\geq 2\sigma$ redder than this fit. Fitting for the line- and \pz-selected samples were performed separately, and we show the best fit (solid black line) and $2\sigma$ selection criteria (dashed line) for the line-selected sample only. The best fit and selection for the \pz-selected sample is similar to the line-selected. The side panels show four examples of different kinds of SEDs found in this diagram, including a blue galaxy with very strong emission lines (bottom right), an average ELG with a flat continuum and strong emission lines (bottom left), galaxy with red continuum fluxes and weak emission lines (top left), and a REG (top right). The EWs reported for each example are measured directly from the medium-band photometry, as described in \S\ref{subsec:measuring_properties}.}
    \label{fig:ColourColour_Selection}
\end{figure*} 

\subsection{Photometric Redshift Sample}\label{subsec:pz-sample}

The second sample used to search for REGs is selected based on their photometric redshifts, which we refer to as the \pz-selected sample. The \pz-selected sample serves as a reference sample for comparison with the line-selected sample. As discussed above, high-redshift galaxy selections commonly place minimum \snr\ requirements on broad-band filters which biases these samples towards brighter sources. The \pz-selected sample is designed to replicate such samples, by placing minimum \snr\ requirements on the underlying continuum emission of galaxies. As such, objects in the \pz-selected sample are more in-line with existing samples of high-redshift galaxies, making them a useful reference sample to compare to the line-selected sample.

We use a selection similar to that of \cite{willott_steep_2024}, who measured the UV luminosity function for $z > 7.5$ galaxies in CANUCS. In order to keep the \pz-selected sample consistent with the line-selected sample, we extend their selection criteria down to lower redshifts which fully overlap with our line-selected sample. We do this by separating the \pz-selected sample into two redshift bins. For the higher redshift bin, we require:

\begin{enumerate}
    \item $ 7.3 \lesssim z_{phot} \lesssim 9.2$
    \item $z_{160} \geq 6$
    \item $S/N (F277W) \geq 8  $
    \item $S/N(F435W) <  2$
    \item $S/N(F606W) <  2$
    \item $S/N(F090W) <  2$
\end{enumerate}

Here, $z_{phot}$ is the maximum likelihood redshift and $z_{160}$ is the 16th percentile of the redshift probability distribution function from \texttt{EAZY-py}. At these redshifts, $F277W$ measures the rest-frame optical continuum, and the \snr$(F277W)$ requirement ensures sources are well detected at these wavelengths. The additional \snr\ cuts on $F435W$, $F606W$, and $F090W$ rule out low-redshift interlopers by ensuring sources are not detected at wavelengths bluewards of the Lyman break. The cut on $z_{160}$ serves a similar function, by excluding objects which have potential lower redshift solutions. In the lower redshift bin, we adopt similar requirements but adjust the filters to probe lower redshifts:

\begin{enumerate}
    \item $4.9 \lesssim z_{phot} \lesssim 7.3$
    \item $z_{160} \geq 4$
    \item $S/N (F200W) \geq 8$ or $S/N (F182M + F210M) \geq 6$
    \item $S/N(F435W) <  2$
    \item $S/N(F606W) <  2$
\end{enumerate}

In this case, the \snr\ requirements on $F182M$, $F210M$, and $F200W$ serve the same purpose as the \snr$(F277W)$ requirement in the high redshift bin. We utilize a combination of $F182M + F210M$ or $F200W$ to probe the rest-frame optical continuum, as the filter coverage varies by field. While the three flanking fields with Technicolor observations have coverage in $F182M$, $F210M$, and $F200W$, the two flanking fields without Technicolor observations (MACS0417 and MACS1423) have coverage in only $F182M$ and $F210M$, while the three cluster fields have coverage in only $F200W$. Thus, we use $F200W$ to constrain the rest-frame optical when available, and use $F182M$ and $F210M$ otherwise. We adjust the \snr\ requirements of $F182M$ and $F210M$ to account for the shallower depth of the medium-bands compared to the broad-bands. Following a visual inspection, our \pz\ sample consists of 592 sources across $4.89 < z < 8.89$. 

\subsection{LRD Sample} \label{subsec:LRD_Selections}

In addition to the line-selected and \pz-selected samples, we search for LRDs in our dataset to cross-check with the sample of REGs. We apply the LRD selections of \citet{kokorev_census_2024} and \cite{kocevski_rise_2025}, which yields a sample of 75 LRDs over $1.16 < z < 9.08$. Most LRDs meet both selection criteria, while a handful are only selected by one. We include all LRDs in our sample for completeness.

In order to accurately compare the LRDs with the line- and \pz-selected samples, we perform several additional cuts to ensure the samples are consistent. First, we limit the sample of LRDs to $ 4.89 < z < 9.23$ to align with the line- and \pz-selected samples. Additionally, we require the LRDs have observations in HST bands. We highlight that neither \citet{kokorev_census_2024} or \cite{kocevski_rise_2025} impose such a requirement. However, we choose to include this in our LRD selection for consistency, as both the line- and \pz-selected samples require sources have observations in bands blue-wards of the Lyman break. At $ z > 4.89$, this requires observations in at least one HST band (e.g. $F435W$ and $F606W$). Thus we exclude LRDs which lack the relevant HST observations. The \pz\ cut removes 31 LRDs from the sample, while the HST coverage requirement removes 24 LRDs. 

\begin{figure*}
    \centering
    \includegraphics[width=0.9 \linewidth]{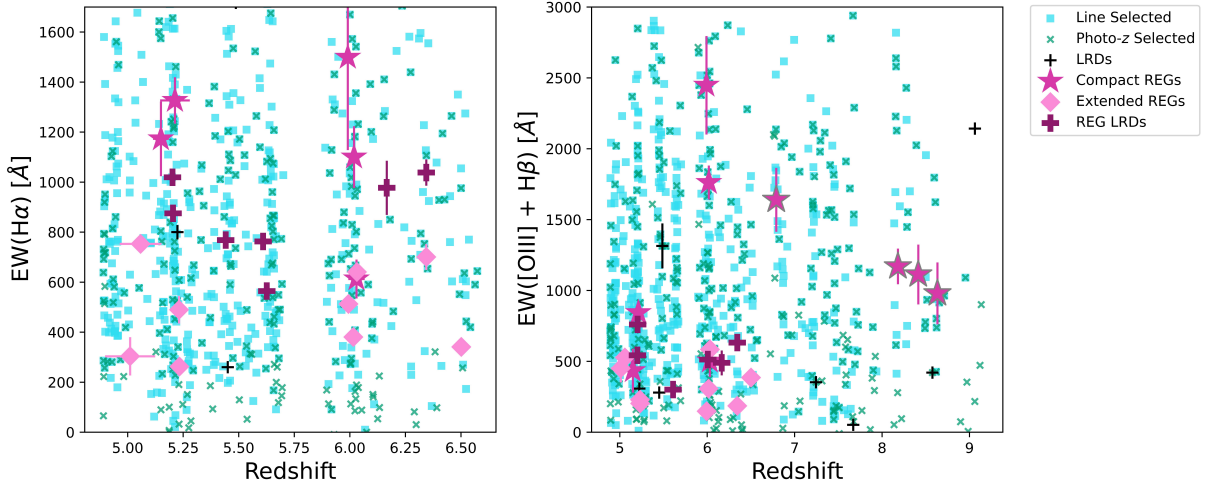}
    \caption{EW(\Ha) (left) and EW(\OIIIHb) (right) vs. photometric redshift for our samples. REGs which ony heave measurements of the \OIIIHb\ emission lines are outlined with a grey outline. The sample of REGs have a wide range of EWs, and include both some of the lowest and highest EW objects in the line- and \pz-selected samples.}
    \label{fig:EW_z}
\end{figure*}

Finally, we perform an additional redshift cut on LRDs found in the MACS0417 and MACS1423 flanking fields. These fields have CANUCS observations in $F410M$, but lack the additional $F430M$, $F460M$, and $F480M$ imaging provided by Technicolor and JUMPS. As discussed in \S\ref{subsec:measuring_properties} and \S\ref{sec:reg_selection}, our emission line measurements (e.g. EW) and REG selections require that one of the \Ha\ or \OIIIHb\ emission lines falls in a $\sim 4\mu$m medium-band. As such, we are unable to measure the emission line properties of objects in MACS0417 and MACS1423 which have the \Ha\ or \OIIIHb\ emission lines redwards of $F410M$. While such objects can still be selected as LRDs based on the selection criteria of \citet{kokorev_census_2024} and \cite{kocevski_rise_2025}, they will be excluded from our sample of REGs because they lack $F430M$, $F460M$, and $F480M$ observations. Thus we exclude them from this analysis to ensure consistency. This cut excludes an additional 5 LRDs, bringing our sample size to 15 LRDs.

\subsection{Physical and Spectral Properties} \label{subsec:measuring_properties}

We measure the \Ha\ and \OIIIHb\ EWs for all three samples defined above using the methods outlined in \cite{vilella-rojo_extracting_2015}. Since the line-selected sample contains a combination of objects with bright, well detected continuum fluxes and faint, undetected continuum fluxes, we employ several different methods to measure the underlying continuum. The first method is applicable to bright galaxies which have continuum detections in several filters. In this case, we fit a power law of the form $f_\lambda \propto \lambda ^\beta$ to the underlying continuum using filters which are uncontaminated by prominent spectral features, such as strong emission lines, the Lyman break, and the Balmer break. Thanks to the extensive medium-band coverage in these fields, there are $\sim 9 - 11$ bands available to fit the continuum at each redshift, of which we require at least three to have $S/N \geq 3$. We then fit these filters with a power law, and use this fit to measure the continuum in medium-bands containing the \Ha\ or \OIIIHb\ emission lines.

If fewer than 3 filters have $S/N \geq 3$, we measure the continuum using the uncontaminated medium-band filter(s) closest in wavelength to the filter containing an emission line. If there are two neighbouring bands with $S/N \geq 3$ available (one bluewards and one redwards of the band with the emission line), we take the average of those two bands and assume they are representative of the underlying continuum. Alternatively, if only one neighbouring band with $S/N \geq 3$ is available, we assume a flat continuum and take the flux in that band to be representative of the underlying continuum at nearby wavelengths. Finally, if there are no neighbouring bands with $S/N \geq 3$, we take the $2\sigma$ limit of the neighbouring band with the highest \snr\ as a continuum measurement, and calculate a lower limit on the EW.

To estimate the physical sizes, we use \texttt{GALFIT} (\citealt{peng_detailed_2010}) to fit single-component Sersic profiles to our sources. Among the parameters fit are the effective radius (\reff), Sersic index ($n$), axis ratio ($b/a$), and total magnitude. The effective radius was allowed to vary between $0.04'' < R_{eff} < 4''$, Sersic index between $0.2 < n < 10$, axis ratio between $0.05 < a/b < 1$, and total magnitude within $\pm3$ magnitudes of the photometry. Additionally, we convert the fitted effective radii to circularized radii and measure the circularized effective radius in kiloparsecs (kpc). Full details of the fitting will be described in Merchant et al. in prep. 

Finally, we perform additional SED fitting using \texttt{BAGPIPES} (\citealt{carnall_inferring_2018}) to estimate the physical properties of the line and \pz-selected samples. While AGN templates are included in \texttt{BAGPIPES}, we exclude the LRDs from this analysis as there is no generally accepted method of measuring their physical properties using \texttt{BAGPIPES}. The \texttt{BAGPIPES} fitting is performed within $z_{EAZY-py}\pm 0.05$ of the photometric redshift, assuming a double power-law star formation history, \cite{chabrier_galactic_2003} IMF, and \cite{calzetti_dust_2000} dust law. We utilize the same priors as those in the CANUCS data release (outlined in Table 4 of \citealt{sarrouh_canucstechnicolor_2026}), however we re-do the fitting in this work to utilize the JUMPS observations, and to ensure the photometric redshifts used in the SED fitting agree with those from \texttt{EAZY-py}. As our sources are at high-redshift, we fit the \texttt{COLOR03} apertures, as this ensures the highest \snr\ for our sources.

\begin{figure*}
    \centering
    \includegraphics[width=0.9   \linewidth]{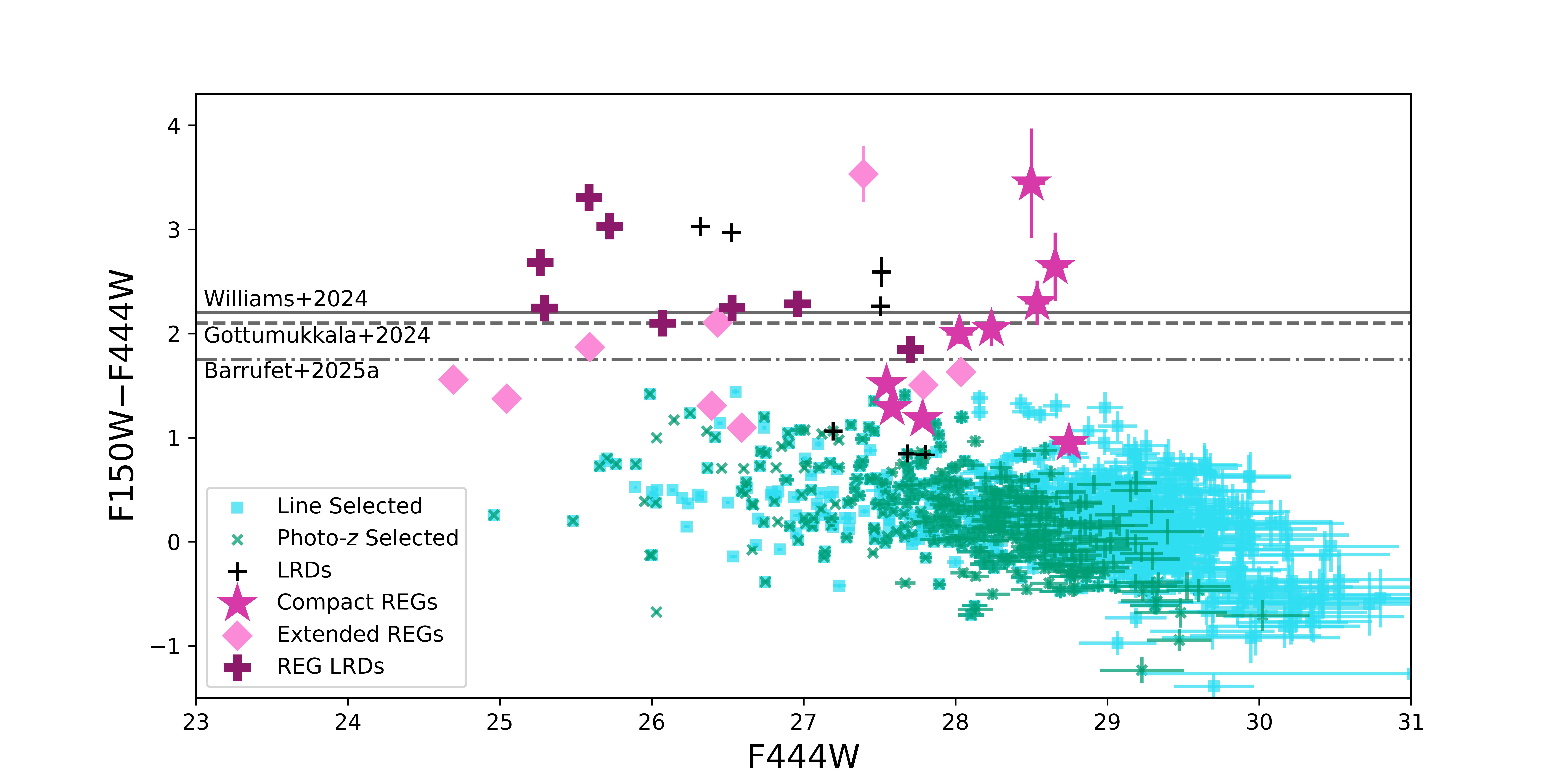}
    \caption{$F150W - F444W$ vs. $F444W$ magnitude diagram for our samples. We also show the $F150W - F444W$ ERO color cuts from \citealt{williams_galaxies_2024} (solid line), \citealt{gottumukkala_unveiling_2024} (dahsed line), and \citealt{barrufet_quiescent_2025} (dash-dot line). Roughly half the REGs qualify as EROs based on at least one of these criteria, and the REG LRDs have similar colors and magnitudes to the full sample of LRDs in CANUCS. }
    \label{fig:ERO_Selection}
\end{figure*}
\section{A Population of Red Galaxies with Strong Emission Lines} \label{sec:reg_selection}
\subsection{Correlation Between Continuum Colors and Line Emission Strength}\label{subsec:correlation}

We investigate the relationship between galaxy continuum colors and emission line strengths using four color-color diagrams in Figure \ref{fig:ColourColour_byz}. The colors in each panel of Figure \ref{fig:ColourColour_byz} are chosen to highlight two features of galaxy SEDs. The first color is chosen to measure the strength of the \Ha\ or \OIIIHb\ emission lines ($x-$axis). This emission line color is measured using two nearby medium-bands, including one which contains the \Ha\ or \OIIIHb\ emission lines (one of the $F410M$, $F430M$, $F460M$, or $F480M$). The second medium-band is a neighbouring filter free of strong emission lines, which provides an estimate of the underlying continuum flux nearby the \Ha\ or \OIIIHb\ emission lines (either $F360M$ or $F410M$). This ``continuum band'' is always bluewards of and close in wavelength to the filter containing the emission line. In order to ensure that these colors provide accurate measurements of emission line strengths, we perform a redshift cut on the line and \pz\ selected samples. This cut ensures that one of the \Ha\ or \OIIIHb\ emission lines is fully contained within the correct medium-band filter, and that the continuum band is not contaminated with nearby emission-lines. The redshift cuts and continuum bands used in each panel are summarized in Table \ref{tab:Selection_Summary}, and are highlighted in each panel of Figure \ref{fig:ColourColour_byz}. 

The second feature highlighted in Figure \ref{fig:ColourColour_byz} is the rest-frame UV-optical continuum ($y-$axis). This is probed using the $F150W$ filter, which measures the rest-frame UV continuum, paired with the continuum band described above, which measures the rest-frame optical continuum. In order to ensure accurate measurements of the optical continuum, we perform an \snr\ cut of $S/N\geq 3$ on the band that measures the optical continuum. This is necessary since the line-selected sample is designed to identify faint ELGs, and contains many ``emission-line only'' sources that are only well detected in medium-bands that contain strong emission lines. These sources will have low \snr\ in the continuum band, leaving their rest-frame optical continua poorly constrained. As such, these sources are unsuitable for our analysis. However, we do not perform any \snr\ cut on $F150W$ since many red galaxies are faint or undetected in the rest-frame UV. In total, this produces a sample of 862 line-selected galaxies and 405 \pz\ selected galaxies over $4.9 < z < 8.9$.

Each color-color diagram in Figure \ref{fig:ColourColour_byz} has a large diversity of galaxy colors. These range from strong line emission (very red $x-$axis colors) and blue continuum colors, to weaker line emission (bluer $x-$axis colors) and redder continuum colors. We note that the subset of line-selected galaxies with line emission colors $\sim 0$ mag are sources which have \Ha\ in the $\sim 4\mu$m medium-bands. All these sources were selected based on their \OIIIHb\ emission lines which fall in medium-bands bluewards of $F410M$, and thus their \OIIIHb\ emission line colors are not included in Figure \ref{fig:ColourColour_byz}. These sources have generally weak \OIIIHb\ lines (EW$\lesssim 500$\AA) and even weaker \Ha\ emission which leads to line emission colors $\sim 0 $ in Figure \ref{fig:ColourColour_byz}. Including such sources in our analysis allows us to assess the effect of low-EW emission lines to the correlation.

In each panel of Figure \ref{fig:ColourColour_byz}, there is an apparent correlation between continuum colors and emission line strength, where galaxies with bluer continuum colors tend to have stronger emission lines (bottom right), and galaxies with redder continuum colors have weaker emission lines (top left). To further investigate this correlation, we combine each panel of Figure \ref{fig:ColourColour_byz} onto a single panel in Figure \ref{fig:ColourColour_Selection}. Here, we use the same colors as in Figure \ref{fig:ColourColour_byz}, but they are re-labelled as simply a line color ($x-$axis) and continuum color ($y$-axis). When combining the four panels of Figure \ref{fig:ColourColour_byz}, we remove any duplicate objects which appear in more than one panel. These duplicates occur when one emission line appears in two overlapping medium-bands (either $F410M$ and $F430M$ or $F460M$ and $F480M$). We exclude the filter with the weaker line emission color ($x$-axis), and include only the stronger color in Figure \ref{fig:ColourColour_Selection}. In all cases, the continuum color ($y$-axis) is unaffected by which filter is used to measure the line emission. Additionally, we note that none of the objects in Figure \ref{fig:ColourColour_byz} appear in two panels due to the presence of \Ha\ and \OIIIHb\ in two different filters. This is a consequence of the \OIIIHb\ lines only entering $F410M$ at $z \approx 6.7$, by which point \Ha\ has been redshifted out of $F480M$.

In order to test if there is a correlation between emission line strength and continuum colors in Figure \ref{fig:ColourColour_Selection}, we perform a Spearman's correlation test on the line and \pz\ selected samples. The Spearman's correlation tests yields coefficients of $\rho_\text{line} = -0.34$ and $ \rho_\text{\pz} = -0.38$, with $p$-values of $p_\text{line} \sim 1\times 10^{-25}$ and $p_\text{\pz} \sim 1\times 10^{-13}$, respectively. This confirms the correlation between emission line strength and continuum color, albeit with a large scatter. Performing the Spearman's test on each of the panels in Figure \ref{fig:ColourColour_byz} individually yields similar results, with a statistically signification correlation in each panel.

In order to assess the influence of the filters used to measure continuum colors on this correlation, we perform the Spearman's correlation test on two variants of Figure \ref{fig:ColourColour_Selection}. The first variant tests if the correlation is driven by our choice to use the same medium-band to measure the rest-frame optical continuum on both axes, as this can produce an inherent correlation in low \snr\ data. We test this by changing which medium-band filters are used to measure the rest-frame UV-optical continuum colors: for sources with emission lines $F410M$ or $F430M$ we measure the UV-optical continuum using an $F150W - F335M$ color (instead of $F150W - F360M$), while we use an $F150W - F360M$ color for sources with emission lines in $F460M$ or $F480M$ (instead of$F150W - F410M$). We leave the filters used to measure the line emission colors unchanged so that the rest-frame optical is being probed by different filters on each axis. Performing a Spearman's correlation test on the resulting distribution produces a coefficient of $\rho = -0.14$ and $p \sim 4 \times 10^{-3}$. This confirms a statistically significant correlation between emission line strength and continuum color, although weaker than the distribution in Figure \ref{fig:ColourColour_Selection}. 

Additionally, we assess the effect of using the same bands to measure the rest-frame UV ($F150W$) and optical ($F360M$ or $F410M$) continuum flux, irrespective of redshift. For galaxies in the lower-redshift range of our sample where \Ha\ falls in one of the $\sim 4 \mu$m medium-bands, $F150W$ samples rest-frame wavelengths of $0.19 \lesssim \lambda_{\text{rest, UV}} \lesssim 0.25$ while $F360M$ and $F410M$ samples rest-frame wavelengths of $0.54 \lesssim \lambda_{\text{rest, Opt}} \lesssim 0.61$. However, in the higher-redshift range where \OIIIHb\ falls in one of the $\sim 4 \mu$m medium-bands, the rest-frame wavelengths probed by these filters shift to $0.14 \lesssim \lambda_{\text{rest, UV}} \lesssim 0.19$ in the UV and $0.40 \lesssim \lambda_{\text{rest, Opt}} \lesssim 0.47$ in the optical. 

In order to test the effect this has on the correlation, we classify our sample into two sub-samples based on which emission line falls in the $\sim 4\mu$m medium-bands: an \Ha\ sub-sample covering $4.89 \leq z \leq 6.58$, and an \OIIIHb\ sub-sample covering $6.71 \leq z \leq 9.23$. Performing the Spearman's correlation test on each sub-sample confirms that both have statistically significant correlations between continuum color and emission line strength, producing $\rho_{\text{H}\alpha} = -0.41$ and $p_{\text{H}\alpha} = 1 \times 10^{-26}$ for the \Ha\ sub-sample and $\rho_{\text{[OIII] + H}\beta} = -0.19$ and $p_{\text{[OIII] + H}\beta} \sim 6 \times 10^{-3}$ in the \OIIIHb\ sub-sample. While this test does confirm that the correlation in Figure \ref{fig:ColourColour_Selection} is real, it also highlights a redshift dependence where the strength of the relation weakens at higher-redshifts. While we cannot rule out that this effect is real, it is likely a selection effect where the continuum colors are probing a smaller range of rest-frame wavelengths at higher redshifts, leading to a weaker correlation at high-redshift than at low-redshift. Such behaviour could be accounted for by changing which continuum bands are used as a function of redshift so the medium-bands always sample the same rest-frame wavelengths. However, this is challenging owing to different filter depths and coverage in the fields used in this work. Thus, given the evidence that this correlation is real across the full redshift range we investigate, we do not account for this effect.

Qualitatively, the correlation between continuum color and emission line strength makes sense. The most powerful ELGs will populate the bottom right of Figure \ref{fig:ColourColour_Selection}, with intrinsically blue continuum colors due to the young stellar populations dominating their SEDs. These stars also power high EW \Ha\ and \OIIIHb\ emission lines, in some cases reaching EW $> 2000$\AA. As the most massive stars evolve off the main sequence, the underlying continuum fluxes will flatten and eventually redden, and emission lines will weaken. This will push galaxy colors towards the centre of Figure \ref{fig:ColourColour_Selection}, with flatter continuum colors and more moderate emission lines (EWs $\sim 1000$\AA). Galaxies will then be driven to the top left of Figure \ref{fig:ColourColour_Selection} with the reddest continuum fluxes and weakest emission lines (EWs $\lesssim 500$\AA).

The presence of dust in galaxies has a similar effect by reddening the continuum and obscuring emission lines. The nebular regions where emission lines originate have been shown to experience more dust attenuation than the stellar continuum by a factor of $\sim2\times$ (e.g. \citealt{calzetti_dust_1994}, see \citealt{salim_dust_2020} for a review), which lowers EWs and pushes galaxies towards the top right of Figure \ref{fig:ColourColour_Selection}. The side panels of Figure \ref{fig:ColourColour_Selection} have examples of each type of SED discussed above. 

Notably, there is a small population of galaxies occupying the top right of the color-color diagrams in Figures \ref{fig:ColourColour_byz} and \ref{fig:ColourColour_Selection}. Seemingly at odds with the correlation discussed above, these objects have a combination of high EW \Ha\ and \OIIIHb\ emission lines and red continuum fluxes. We refer to this population of galaxies as the Red Emission line Galaxies (REGs).

\begin{table}[]
    \centering
    \begin{tabular}{c|c|c}
        Flux   & Redshifts  & Continuum \\
        \noalign{\vskip -3pt}
        Excess &        Selected            & Band \\
        \hline 
        \hline 
         $F410M$ & H$\alpha$: $4.89 \leq z \leq 5.56 $         & $F360M$ \\
               & [OIII] + H$\beta$: $6.72 \leq z \leq 7.85 $ &  \\
        \hline
         $F430M$ & H$\alpha$: $5.35 \leq z \leq 5.70 $         & $F360M$ \\
               & [OIII] + H$\beta$: $7.32 \leq z \leq 8.05 $ &  \\
        \hline 
         $F460M$ & H$\alpha$: $ 5.88 \leq z \leq 6.32 $         & $F410M$ \\
               & [OIII] + H$\beta$: $8.02 \leq z \leq 8.77 $ &  \\
        \hline 
         $F480M$ & H$\alpha$: $ 6.10 \leq z \leq 6.58 $         & $F410M$ \\
               & [OIII] + H$\beta$: $ 8.31 \leq z \leq 9.23 $ &  \\
    \end{tabular}
    \caption{Summary of which filters are used to create the color-color diagrams in Figures \ref{fig:ColourColour_byz} and \ref{fig:ColourColour_Selection}, including the redshifts targeted. }
    \label{tab:Selection_Summary}
\end{table}

\begin{figure*}
    \centering
    \includegraphics[width=0.9\linewidth]{real_final_figures/Av_compactness_JUMPS_v7.png}
    \caption{$A_V$ (left), EW(\Ha) (middle) and EW(\OIIIHb) (right) vs. \reff\ measured using \texttt{GALFIT} in $F444W$ of the REGs and LRDs in CANUCS (defined in \S\ref{subsec:LRD_Selections}). \reff\ is reported in arcseconds, with the vertical dashed line showing the FWHM of the $F444W$ PSF. Nine of the REGs are resolved in $F444W$ with effective radii larger than the $F444W$ PSF, while remaining REGs are unresolved and have \reff\ consistent with zero within uncertainties. Eight of these unresolved REGs qualify as LRDs based on the selection criteria \citealt{kokorev_census_2024} and \citealt{kocevski_rise_2025} (see \ref{subsec:LRD_Selections}), which we refer to as the REG LRDs. We refer to the remaining nine unresolved REGs as the compact REGs. Objects with black outlines are experiencing lensing magnification, with $\mu \sim 1 - 2$. We report the lensing corrected \reff.}
    \label{fig:AV_compactness}
\end{figure*}

\subsection{Sample of Red Emission Line Galaxies}\label{subsec:REGs_sample}

To systematically identify REGs in our samples, we fit straight lines to both the line and \pz\ selected samples in Figure \ref{fig:ColourColour_Selection}. Both the line and \pz\ selected samples have similar fit parameters and similar amounts of scatter. We then use these fits to calculate a typical rest-frame UV-optical continuum color as a function of the strength of the \Ha\ and \OIIIHb\ emission lines. We also measure the scatter, $\sigma$, of the relation. The REGs are then defined as galaxies whose continuum colors are $ \geq 2 \sigma$ redder than the typical continuum color based on the fit ($\geq 0.82$ mag and $\geq 0.90$ mag for the line and \pz\ selected samples, respectively). We show the best fit (solid black line) and color selection (dashed line) for the line-selected sample in Figure \ref{fig:ColourColour_Selection}.

This produces a sample of 29 objects across $5.01 < z < 8.63$, 26 of which are in the line-selected sample and 12 of which are in the \pz-selected sample, with 9 being found in both samples. The 3 objects found uniquely in the \pz-selected sample all have very weak emission lines, with line emission colors $< 0.3$ mag and EWs $\lesssim 100$\AA. Thus, while these 3 objects qualify as REGs in principle, they do not have emission lines strong enough to qualify as REGs in practice, and we exclude them from our sample. The REGs are highlighted as pink stars in Figure \ref{fig:ColourColour_Selection}, with REGs selected based on their \OIIIHb\ emission lines (4 in total) outlined in black and REGs selected on their \Ha\ emission (22 in total) have no outlines.

\begin{figure*}
    \centering
    \includegraphics[width=\textwidth]{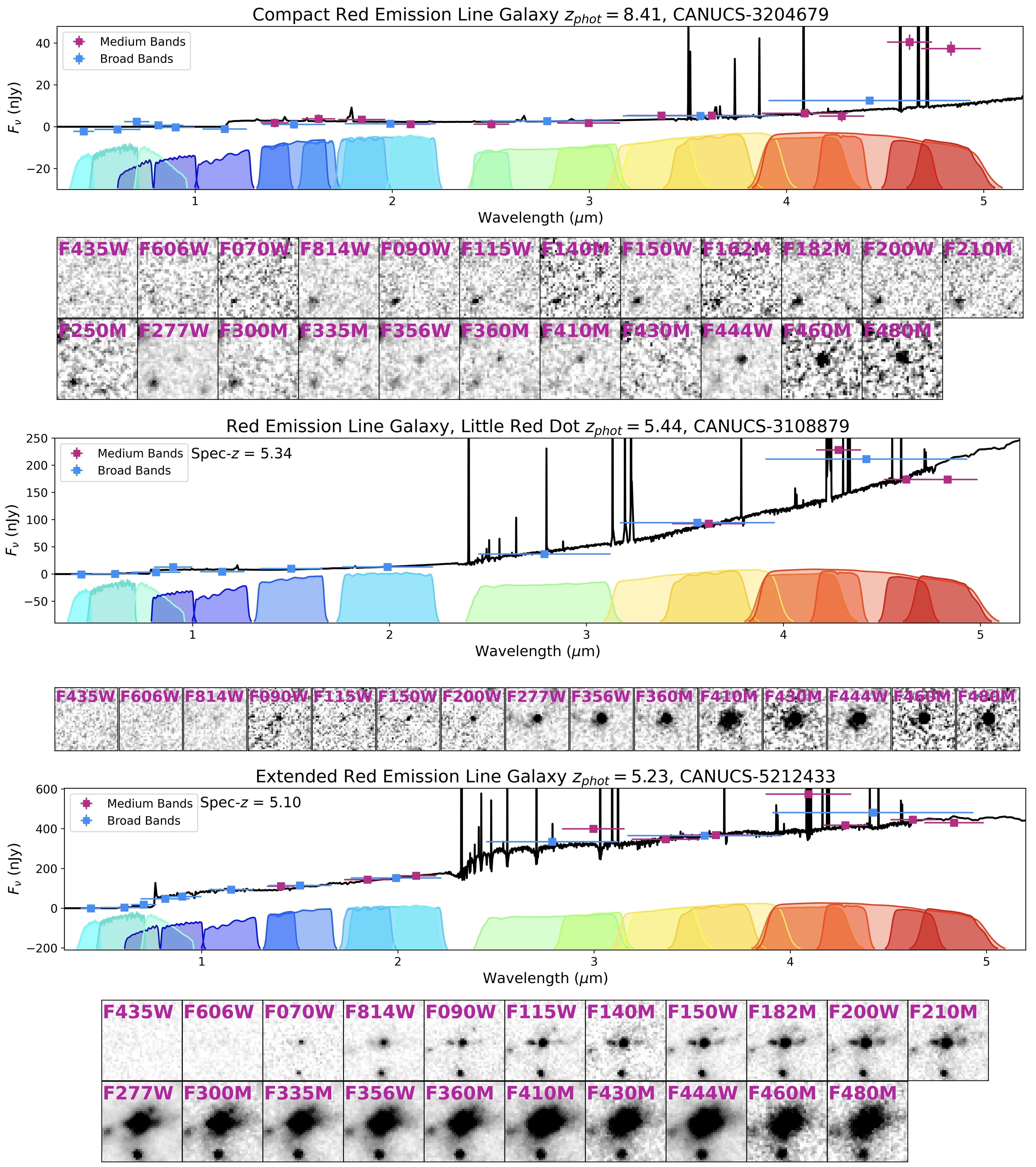}
    \caption{Three examples of very Red Emission line Galaxies with the \texttt{EAZY-py} SED fits, photometry, and cutouts. This includes examples of all three types of REGs (\S \ref{sec:Nature_REGs}): compact, LRD, and extended REGs are shown in the top, middle and bottom panels, respectively.}
    \label{fig:examples}
\end{figure*}

We show the EWs of the REGs, line-selected, \pz-selected, and LRD samples in Figure \ref{fig:EW_z} (measured as described in \S\ref{subsec:measuring_properties}). The REGs have a wide range of EWs, and include some of the lowest and highest EW objects in the line-selected sample. All 22 REGs which have their \Ha\ emission-lines observed by the medium-bands qualify as ELGs based on their \Ha\ emission, with EW(\Ha) $\gtrsim 500$\AA. However, there is a wider variation in the EW(\OIIIHb) of the REGs. Of the 24 REGs which have their \OIIIHb\ emission lines constrained by the medium-bands, there are several which have very low EW(\OIIIHb), with some as low as $\sim 150$\AA. Such objects qualify as REGs based on their higher EW(\Ha) emission, but have much weaker \OIIIHb. Conversely, several of the REGs have extremely high EW(\OIIIHb) emission lines, up to EW(\OIIIHb)$\sim 2300$\AA. 

Given that the REGs are selected based on red continuum colors, it is useful to compare them with existing samples of red galaxies from the literature. In Figure \ref{fig:ERO_Selection} we show an $F150W - F444W$ vs. $F444W$ color-magnitude diagram for our line-selected, \pz\ selected, and REG samples, along with three color cuts used to identify EROs from \citealt{williams_galaxies_2024} ($F150W - F444W > 2.2$ mag), \citealt{gottumukkala_unveiling_2024} ($F150W - F444W > 2.1$ mag), and \citealt{barrufet_quiescent_2025} ($F150W - F444W > 1.75$ mag). The REGs exhibit the reddest $F150W - F444W$ colors among the line and \pz\ selected samples, with colors ranging from $F150W - F444W \sim 1$ mag to $F150W - F444W \sim 3.5$ mag. Roughly $50\%$ of the REGs (16/26) satisfy at least one of the ERO selections based on $F150W - F444W$ color.

Although many of our REGs qualify as EROs, there are important differences between the samples. Both REG and ERO selections utilize $1.5\mu$m $- \sim 4\mu$m color selections. However, ERO selections typically use a single $F150W - F444W$ or $F150W - F356W$ broad-band color cut while our selection is more complex, involving two color cuts and multiple medium-band filters. This allows us to more robustly identify galaxies with red continuum colors, and more accurately measure their properties, including emission line EWs, continuum colors, and redshifts than typical samples of EROs. Additionally, our choice to search for REGs from an already robust sample of ELGs (the line-selected sample) allows us to efficiently identify REGs with faint $F444W$ magnitudes. While the line-selected sample will exclude many red galaxies without strong emission lines, identifying REGs in a sample of medium-band selected line emitters allows us to extend our sample down to $> 28$ mag in $F444W$. Indeed, 7 of the REGs have $F444W > 28$mag, all of which are found only in the line-selected sample. Such objects would be excluded from many ERO selections on the basis of their faint $F444W$ magnitudes, and thus remain poorly understood (e.g. \citealt{williams_galaxies_2024}).

\section{The Nature of Red Emission Line Galaxies} \label{sec:Nature_REGs}

In this section, we classify the REGs into three categories based on their colors and morphologies (LRD, extended REGs, and compact REGs), and explore potential origins of their spectral properties.

\subsection{Little Red Dots} \label{subsec:LRDs}

Inspecting the morphologies of the REGs, we can divide the sample into sources with compact and extended morphologies. This classification is highlighted in Figure \ref{fig:AV_compactness}, which shows $A_V$ (left), EW(\Ha) (middle), and EW(\OIIIHb) (right) vs. effective radii (\reff) in arcseconds measured in $F444W$ (see \S \ref{subsec:measuring_properties}). The compact sources are unresolved in $F444W$ with \reff\ consistent with point sources within uncertainties. We take the FWHM of the $F444W$ PSF ($0.161''$, \citealt{sarrouh_canucstechnicolor_2026}) as an upper limit on the sizes, which translate to $R_{eff} \lesssim 1$ kpc at $z \approx 5.2$ and $R_{eff} \lesssim 0.75$ kpc at $z \approx 8.6$. The remaining sources are all resolved in $F444W$. 

Given the red optical colors, compact morphologies, and strong emission lines often observed in LRDs (e.g. \citealt{hviding_rubies_2025}, \citealt{setton_little_2025}), they provide a natural explanation for the REGs with compact morphologies. As such, we compare our sample of REGs to the sample of LRDs identified in \S\ref{subsec:LRD_Selections} and find that eight of the REGs are LRDs based on the selection criteria of \citealt{kokorev_census_2024} and \citealt{kocevski_rise_2025}. We refer to the REGs classified as LRDs as the REG LRDs, the remaining REGs with compact morphologies as the compact REGs (nine in total), and the REGs with extended morphologies as the extended REGs (nine in total). We note that three of the REG LRDs and two of the compact REGs are experiencing lensing magnification. The two lensed compact REGs, CANUCS-3112462 and CANUCS-2113181, have $\mu = 1.70$ and $\mu = 2.14$, respectively, and remain unresolved even with lensing magnification. We highlight lensed sources in Figure \ref{fig:AV_compactness} using black outlines, and report their magnification corrected effective radii. 

\begin{figure*}
    \centering
    \includegraphics[width=0.9   \linewidth]{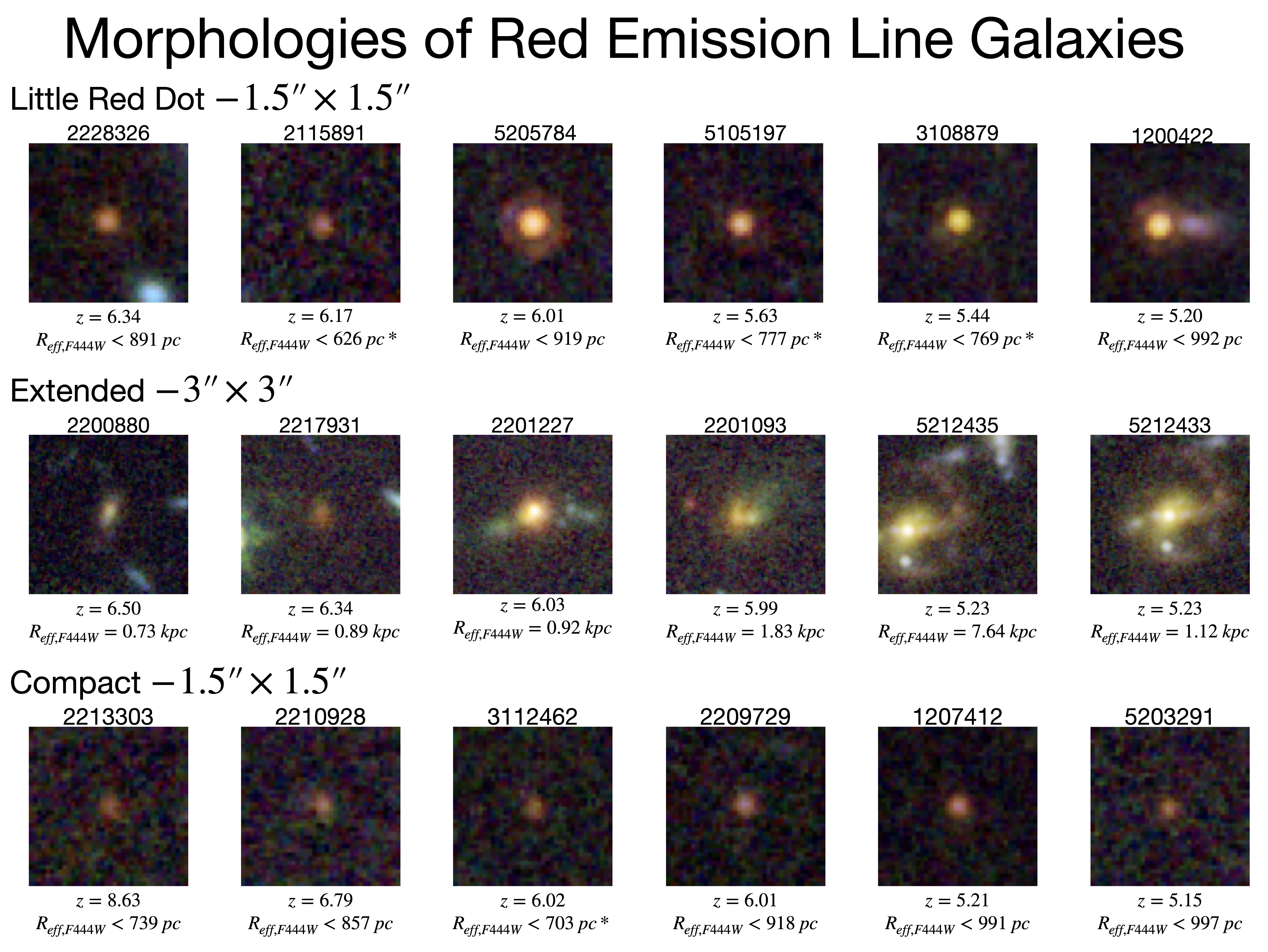}
    \caption{Six cutouts for each type of REG, including REG LRDs (top row) extended REGs (middle row) and compact REGs (bottom row). The cutouts are created using the same filters used to create the color-color diagrams in Figures \ref{fig:ColourColour_byz} and \ref{fig:ColourColour_Selection}, $F150W$ (blue), the continuum band (green), and the band containing either the \Ha\ or \OIIIHb\ emission lines (red, summarized in Table \ref{tab:Selection_Summary}). We report the photometric redshifts of each object, along with the \reff (in kpc) in $F444W$ from \texttt{GALFIT} for the extended REGs. For the REG LRDs and compact REGs, we report the upper limit on \reff\ (in pc) measured from the FWHM of the $F444W$ PSF, $0.161''$. Objects marked with an asterisk are experiencing lensing magnification, with $\mu \sim 1 - 2$. We correct the upper limits on \reff\ for the lensing when applicable.}
    \label{fig:morphologies}
\end{figure*}

Figure \ref{fig:examples} shows the SEDs and cutouts for one of each type of REG, and Figure \ref{fig:morphologies} shows postage stamps and effective radii of six REGs of each type. The redshifts, EWs, and $F444W$ magnitudes of all 26 REGs are reported in Table \ref{tab:properties}. The effective radii of the extended REGs reported in Figure \ref{fig:AV_compactness} are the values measured from \texttt{GALFIT} (see \S\ref{subsec:measuring_properties}), while we report upper limits on the effective radii of the compact REGs and REG LRDs measured from the $F444W$ PSF in Figure \ref{fig:morphologies}. We note that several of the REGs have poor Sersic fits, and thus have unreliable \reff\ measurements. This is likely caused by the relatively low \snr\ in $F444W$ than what is generally required for morphological fitting. However, the difference between the morphologies of the compact and extended REGs is evident through a visual inspection alone: the extended REGs are clearly resolved and often show evidence of structure, whereas the compact REGs appear small and unresolved. Additionally, we incorporate the compactness criteria from \citealt{kokorev_census_2024} and \cite{kocevski_rise_2025} into our LRD selection, which confirms the compact nature of the REG LRDs. Thus, we still classify the REGs into LRDs, compact, and extended categories, despite uncertainties in \reff\ measurements.

The REG LRDs account for roughly half (eight out of 15 LRDs) of the full sample of LRDs (defined in \S\ref{subsec:LRD_Selections}). As highlighted in Figure \ref{fig:ERO_Selection}, the REG LRDs have similar $F150W - F444W$ colors and $F444W$ magnitudes to the full sample of LRDs. However, the EWs of the REG LRDs are, on average, higher than the full sample of the LRDs (Figure \ref{fig:EW_z}). Indeed, most of the seven LRDs which are not classified as REGs are excluded since their emission lines fail to produce sufficiently strong emission-line colors. As such, we conclude that the REG LRDs are part of an abundant population of LRDs which have very red continuum colors and strong emission lines.

\subsection{Extended REGs} \label{subsec:extended_REGs}

\begin{figure*}
    \centering
    \centering
    \includegraphics[width=0.9   \linewidth]{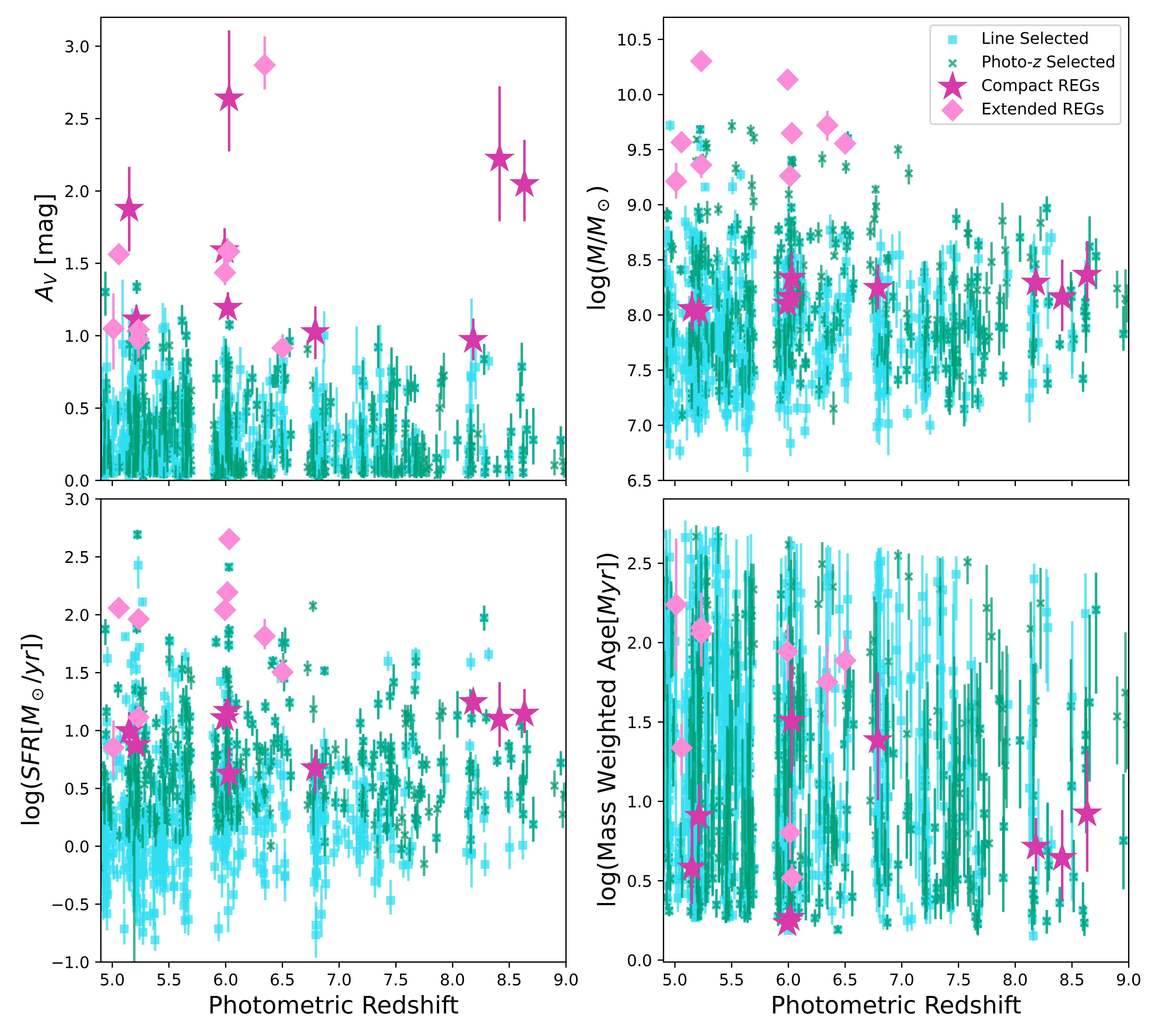}
    
    \caption{The physical properties of our samples, measured from \texttt{BAGPIPES} (excluding objects classified as LRDs). The physical properties of the extended REGs are consistent with DSFGs: high $A_V$ (top left), high stellar masses (top right), high SFR (bottom right), and young mass weighted ages (bottom right). The compact REGs have properties consistent with DSFGs, including high $A_V$. However, they have stellar masses roughly an order of magnitude lower than the extended REGs, and more in-line with the line- and \pz-selected samples. }
    \label{fig:properties}
\end{figure*}

The remaining REGs not classified as LRDs could be explained by dust obscured star-formation or AGN. Here, we explore the possibility that the extended REGs are DSFGs. Figure \ref{fig:properties} shows the physical properties of the REG, line, and \pz\ selected samples from the \texttt{BAGPIPES} SED fits (described in \S \ref{subsec:measuring_properties}), and Table \ref{tab:properties} summarizes these properties. As discussed in \S\ref{subsec:measuring_properties}, we exclude LRDs from Figure \ref{fig:properties} and Table \ref{tab:properties} as there is no universally accepted way to measure their physical properties using \texttt{BAGPIPES}.

The physical properties, colors, and morphologies of the extended REGs are all consistent with those of DSFGs. The extended REGs are among the most dust obscured objects in the line and \pz\ selected samples with average $A_V = 1.45^{+0.04}_{-0.04}$ mag but up to $A_V \sim 3$ mag, and are among the most massive sources at their respective redshifts (average $\log(M_\star/M_\odot) = 10.1^{+0.03}_{-0.03}$). Additionally, they have high SFRs (average $SFR = 115.4^{+7.48}_{-6.08} M_\odot/yr$), and young mass weighted ages (average age $73.8 ^{+33.6}_{-14.69}Myr$). These properties are in-line with DSFGs reported in the literature, and all the extended REGs meet at least one common DSFG selection criteria (e.g. \citealt{alcalde_pampliega_optically_2019}, \citealt{cowie_2_2023}, \citealt{williams_galaxies_2024}, \citealt{mckay_physical_2025},  \citealt{zavala_alma_2025}, \citealt{rodighiero_egs-z11-r0_2026}). 

\begin{figure}
    \centering
    \includegraphics[width=0.8\linewidth]{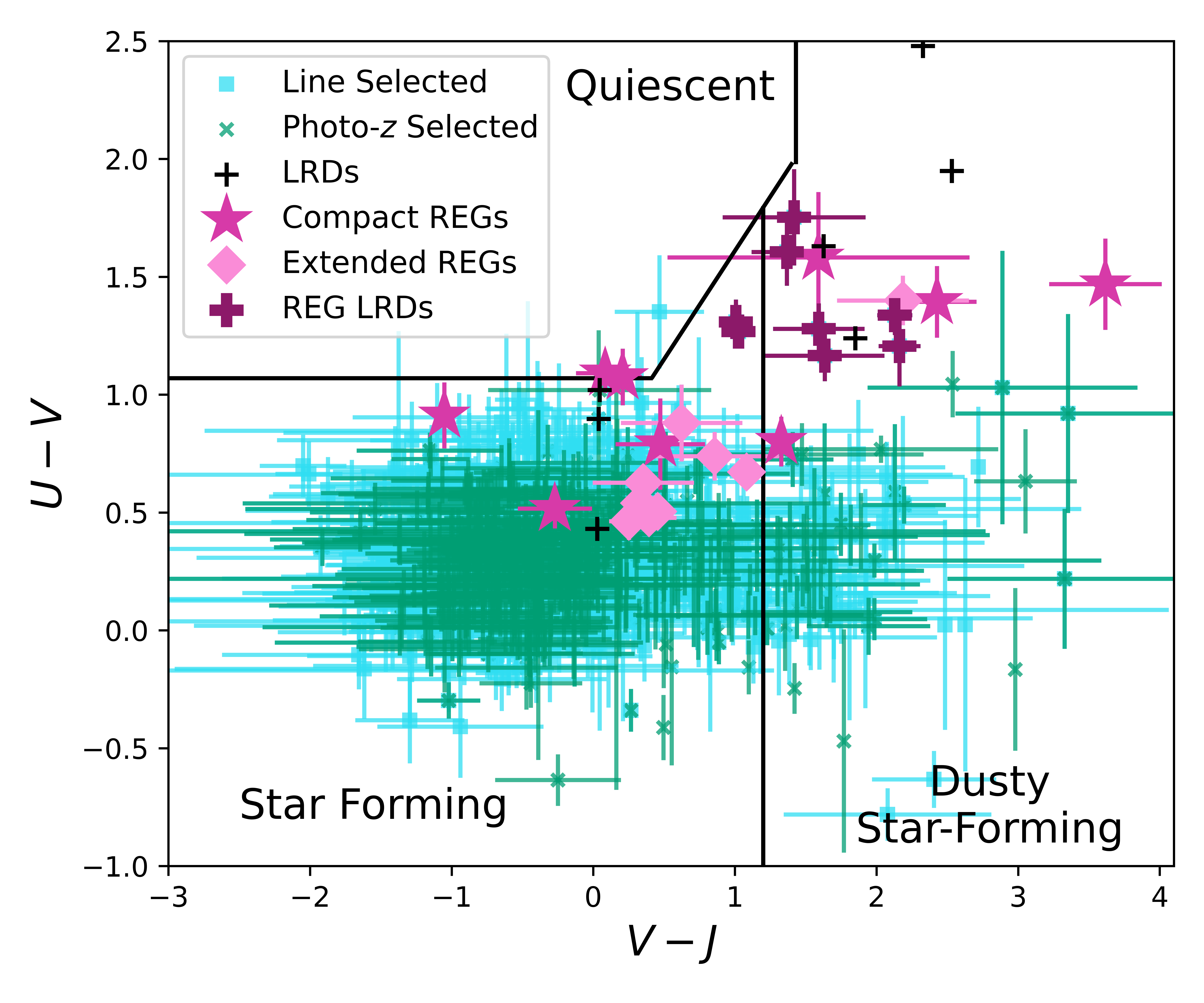}
    \caption{A $UVJ$ diagram for our samples, including the quiescent and star forming regions from \citealt{antwi-danso_beyond_2023}, and the DSFG region from \citealt{spitler_exploring_2014}. All the REGs are classified as star-forming galaxies, but only seven are classified as DSFGs. Additionally, the compact REGs and LRDs have similar $UVJ$ colors.}
    \label{fig:UVJ}
\end{figure}

Figure \ref{fig:UVJ} shows a UVJ diagram for all the galaxy samples discussed in this work, with the $UVJ$ fluxes measured directly from the \texttt{EAZY-py} best-fit SEDs. Figure \ref{fig:UVJ} includes the quiescent and star-forming regions from \cite{antwi-danso_beyond_2023}, as well as the star forming and DSFG regions from \cite{spitler_exploring_2014}. All the extended REGs fall within the star-forming regions, but, notably, only a handful of the extended REGs fall in the \cite{spitler_exploring_2014} DSFG region with the rest falling in the star-forming region. This is likely since the DSFG region is designed to target galaxies with $A_V > 1.6$ mag, and several of the extended REGs fall below that limit. We note that while these sources are far from the most dust extincted sources known at these redshifts (\citealt{sun_identification_2025}, \citealt{bing_almost_2025}), the extended REGs are among the most dust obscured sources in both the line- and \pz-selected samples. Additionally, they have $A_V$ substantially higher than what is typically observed in high-redshift star-forming galaxies (\citealt{roberts-borsani_between_2024}, \citealt{llerena_physical_2024}). Thus, we still consider DSFG a valid interpretation for the extended REGs.

Finally, the morphologies of the extended REGs are also consistent with DSFGs reported in the literature, where there are many examples of DSFGs with similar effective radii ($1 \lesssim R_{eff} \lesssim 3$ kpc, e.g. \citealt{nelson_jwst_2023}, \citealt{mckay_physical_2025}, \citealt{gibson_jades_2024}). Visually inspecting the extended REGs reveals they have complex morphologies, and often have evidence of mergers. Indeed, several of our REGs are located in close proximity to one another. CANUCS-5212433 and CANUCS-5212435 are extended REGs located within $\sim 6$ kpc at $z = 5.23$. CANUCS-5212433 is a massive galaxy ($\log(M_\star/M_\odot) = 10.3$), which appears to be at the centre of a merger with four additional sources within $\lesssim 6 - 7$ kpc of the centre. Similarly, CANUCS-4200621 is a massive, extended REG located at $z = 5.05$ with several sources within $\lesssim 5 - 6$kpc, including another extended REG CANUCS-4222045. Finally, CANUCS-2201093 and CANUCS-2228428 are both located at $z = 5.99$ with a separation of $\sim 6.3$ kpc. CANUCS-2228428 is a compact source and CANUCS-2201093 is an extended, massive galaxy ($\log(M_\star/M_\odot) = 10.16$), which once again appears to be undergoing a merger. Given the physical properties and morphologies of the extended sample, we conclude they are massive DSFGs, many of which are undergoing merger-induced star-formation events.

\begin{figure}
    \centering
    \includegraphics[width=0.8\linewidth]{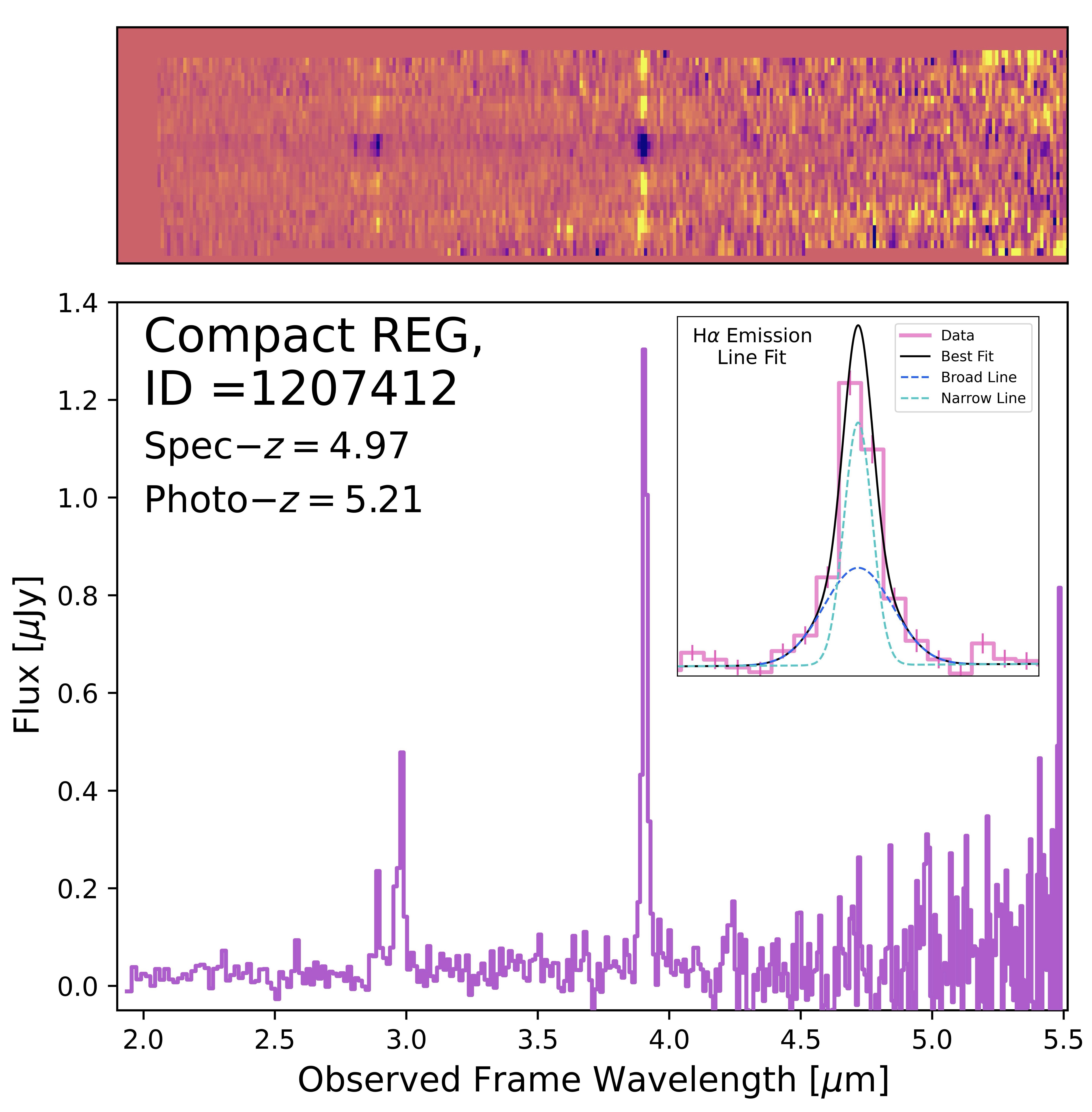}
    \caption{NIRSpec prism spectrum of one of the compact REGs, CANUCS-1207412. The inset shows the double component Gaussian fit to the \Ha\ emission line, including the broad and narrow line components. This fit shows evidence of a broad-line component in \Ha, indicating the presence of an AGN in CANUCS-1207412. The spectrum bluewards of $\lambda_{obs} \sim 2\mu$m falls off the detector edge. }
    \label{fig:nirspec}
\end{figure}

\subsection{Compact REGs as DSFGs} \label{subsec:compact_REGs_DSFGs}

In the following sections, we investigate the nature of the compact REGs, considering dust-obscured star formation or AGN scenarios. At first glance, the compact REGs are consistent with DSFGs. Their physical properties are broadly consistent with DSFGs, including the extended REGs, with high $A_V$ (mean $A_V = 1.63^{+0.1}_{-0.08}$), young mass-weighted ages (mean $10.0^{+6.6}_{-2.55} Myr$), and high Specific Star Formation Rates (SSFRs, mean $\log(SSFR/yr) = -7.16^{+1.2}_{-0.91}$ for the compact REGs and mean $\log(SSFR/yr) = -8.09^{+1.28}_{-1.19}$ for the extended REGs). The compact REGs are also found in similar regions of the $UVJ$ diagram as the extended REGs (Figure \ref{fig:UVJ}), and meet several DSFG selection criteria reported in the literature. While the morphologies of the compact REGs are substantially different to the extended REGs, they are nonetheless consistent with samples of DSFGs reported in the literature, where there are multiple examples of compact/unresolved DSFGs spanning a wide range of redshifts (e.g. \citealt{forrest_magaz3ne_2024} at $z \sim 2 - 3$, \citealt{ganguly_stellar_2025} and \citealt{gomez-guijarro_jwst_2023} at $z \sim 3 - 4$, \citealt{akins_two_2023} at $z \sim 7 - 8$, \citealt{rodighiero_egs-z11-r0_2026} at $z \sim 11$). 

Despite these similarities, there are several properties of the compact REGs that make a DSFG interpretation less straightforward. Firstly, the compact REGs have generally low stellar masses with a mean of $\log(M_\star/M_\odot) = 8.32^{+0.07}_{-0.06}$, roughly an order of magnitude lower than those of the extended REGs. While the lower stellar masses of the compact REGs remain consistent with some samples DSFGs, their substantially lower stellar masses when compared to the extended REGs could indicate fundamental differences between the two samples.

Additionally, the compact REGs have significantly higher EWs than many DSFGs, including the extended REGs. The compact REGs have EWs which are substantially higher than the extended REGs, with average EW(\Ha) $\sim 1135 \pm 187$\AA\ and EW(\OIIIHb) $\sim 1257 \pm 275$\AA\ for the compact REGs and EW(\Ha) $\sim 487 \pm 71$\AA\ and EW(\OIIIHb) $\sim 335 \pm 68$\AA\ for the extended REGs. While there are limited examples of DSFGs with optical emission line strengths comparable to the compact REGs (e.g. \citealt{forrest_magaz3ne_2024} at $z\sim 2 - 3$, \citealt{akins_two_2023} at $z\sim 8$), the high EWs of the compact REGs are difficult to reconcile with the DSFG interpretation when combined with red continuum fluxes and compact sizes of the compact REGs. As with their stellar masses, the difference in EWs between the extended and compact REGs likely suggests that the two samples represent distinct types of DSFGs. In this case, the extended REGs would be comparable to the classical DSFG, while the compact REGs would be a type of lower mass DSFG with high EW emission lines. Conversely, this difference could also imply that the compact and extended REGs have different ionizing sources.

\begin{figure*}
    \centering
    \includegraphics[width = 0.95\linewidth]{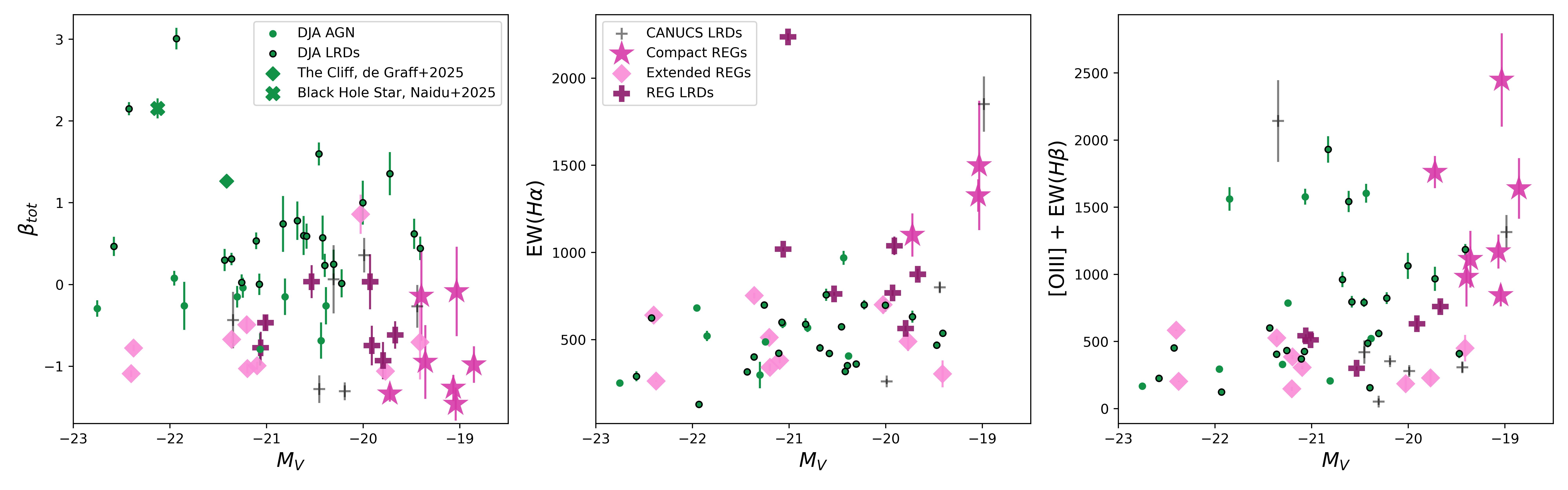}
    \caption{\textbf{Left:} $\beta_{tot}$ vs. $M_V$ for the REGs, a sample of AGN with broad \Ha\ emission lines selected from the DJA (purple circles), including AGN classified as LRDs based on \citealt{hviding_rubies_2025}, and two black hole stars from \citealt{de_graaff_remarkable_2025} and \citealt{naidu_black_2025} (purple diamond and cross). The REGs (particularly the compact REGs) are generally fainter and bluer than the sample of AGN. \textbf{Middle and Right:} EW(\Ha) (middle) and EW(\OIIIHb) (right) for the same the REGs and AGN. The REGs have on average higher EW emission lines than the AGN. }
    \label{fig:AGN}
\end{figure*}

\subsection{Compact REGs as AGN} \label{subsec:AGN}

Finally, we discuss the possibility that the compact REGs are hosting obscured AGN, which would provide a natural explanation for their compact sizes, red continuum fluxes, and strong emission lines. CANUCS NIRSpec follow-up (see \S\ref{sec:observations}) includes prism spectroscopy of three REG LRDs (CANUCS-2115891, CANUCS-5105197, and CANUCS-3108879) and one of the compact REGs (CANUCS-1207412). We show the spectrum of the compact REG (CANUCS-1207412) in Figure \ref{fig:nirspec}. The UV portion of the spectrum for CANUCS-1207412 falls beyond the detector edge, consequently, our observations are limited to the rest-frame optical over $0.3 \mu \text{m} \lesssim \lambda_{rest} \lesssim 0.9\mu \text{m}$. 

Given the broad-line Balmer emission observed in Type I AGN, we search for evidence of a broad-line component CANUCS-1207412 by fitting a double component Gaussian to the \Ha\ emission line, which fits for both narrow- and broad-line components. We measure the Broad-Line Region (BLR) fraction, $f_{BLR}$, defined as the ratio of the line flux in the broad-line component to the total line flux. We find $f_{BLR} = 0.49$ for CANUCS-1207412, indicative of a broad-line component. Additionally, this value of $f_{BLR}$ is in-line with those measured for the three REG LRDs with NIRSpec prism observations, which range from $f_{BLR} = 0.28$ to $f_{BLR} = 0.59$. This value of $f_{BLR}$ suggests the presence of AGN in at least one of the compact REGs. 

Since the rest of the compact REGs lack spectroscopy and other observations commonly used to identify AGN (such as X-ray observations), we cannot conclusively determine if the compact REGs host AGN. Instead, we compare the continuum slopes, absolute magnitudes, and EWs of the compact REGs to a sample of AGN from the Dawn JWST Archive (DJA, \citealt{brammer_dawn_2025}). The sample of AGN were identified from the DJA by searching for sources with NIRSpec prism observations that exhibit broad \Ha\ emission lines at $5 < z < 7.2$. We require \Ha\ \snr\ $>40$, \snr\ $> 20$ in the continuum surrounding \Ha, and \texttt{grade = 3}. As with CANUCS-1207412, we fit the \Ha\ emission line using a double Gaussian, and identify AGN as objects which have evidence for a broad-line region. This produces a sample of 30 AGN, 21 of which are LRDs based on the criteria of \cite{hviding_rubies_2025}. We highlight that LRDs are identified from the sample of AGN based on NIRSpec prism observations, while the CANUCS LRDs are selected using only NIRCam photometry.

In the left panel of Figure \ref{fig:AGN} we show the UV-optical continuum slope, $\beta_{tot}$, vs. the absolute $V-$band magnitude, $M_V$, for all three types of REGs, CANUCS LRDs, and the AGN from the DJA. We measure $\beta_{tot}$ by fitting a power law $f_\lambda \propto \lambda^\beta$ to the continuum over $ 1216$ \AA\ $< \lambda_{rest} < 7000$\AA. For the REGs and CANUCS LRDs, we perform this fitting directly on the photometry using the filters described in \S \ref{subsec:measuring_properties}. For the sample of AGN, we fit this power-law directly to the prism spectra after masking out regions with strong emission lines. We measure $M_V$ by taking the average of a top-hat filter with a width of $100$\AA\ centred on $\lambda_{rest} = 5510$ \AA. We perform this calculation using the \texttt{EAZY-py} SED fit for the REGs and CANUCS LRDs, but measure $M_V$ directly from the prism spectra for the AGN. 

We find very little overlap between the sample of AGN and the compact REGs. The AGN are typically bright, with average $M_V \lesssim -20.9$ mag, while the compact REGs have much fainter $M_V$, on average $M_V\sim -19.0$. Additionally, the compact REGs are bluer than the AGN. The AGN have an average $\beta_{tot} = 0.43$ and the compact REGs have $\beta_{tot} = -0.72$. While this is still a relatively red value for $\beta_{tot}$ (e.g. \citealt{mitsuhashi_discovery_2025}, \citealt{rodighiero_egs-z11-r0_2026}), the compact REGs remain much bluer than the AGN. It is important to note that the difference in $M_V$ between the REGs and AGN may be driven by selection effects, including our requirements for \snr\ $> 20$ in the continuum surrounding \Ha. Nonetheless, our sample of REGs includes several objects with faint $F444W$ magnitudes, with all the compact REGs being fainter than $27$ mag, and six being fainter than $28$ mag (see Figure \ref{fig:ERO_Selection}, Table \ref{tab:properties}). Many selections targeting red galaxies for spectroscopic follow-up apply a magnitude or \snr\ cut on $F444W$ (e.g. $F444W < 27$ mag for RUBIES spectroscopic follow-up, \citealt{de_graaff_rubies_2025}). As a result, similarly faint objects are likely missing from spectroscopically confirmed samples of AGN. 

Additionally, we show the \Ha\ and \OIIIHb\ EWs vs. redshift for the AGN and all three sub-samples of REGs in the middle and right panels of Figure \ref{fig:AGN}. There is more overlap between the EWs of the compact REGs and the AGN, particularly for \OIIIHb, where several AGN have EW(\OIIIHb)$\gtrsim 1000$\AA. However, the compact REGs (average EW(\Ha) $\sim 1142 \pm 183$\AA\ and EW(\OIIIHb) $\sim 1210 \pm 289$\AA) still have significantly stronger emission lines than the samples of AGN (average EW(\Ha)$= 504\pm 42$\AA, EW(\OIIIHb)$= 725\pm 103$\AA). 

The presence of a broad-line component in CANUCS-1207412 and compact morphologies suggest that AGN is a plausible explanation for the compact REGs. If the compact REGs are AGN dominated, they would represent a type of low-luminosity AGN, which have higher EW emission lines than what is generally found in existing samples of AGN. Given the bluer continuum slopes of the compact REGs, they likely qualify as Little Blue Dots (LBDs, \citealt{brazzini_little_2026}, \citealt{asada_origins_2026}), which are Type I AGN that have optical continuum slopes too blue to qualify as LRDs. LBDs, as well as other high-redshift AGN that are not LRDs, remain challenging to identify using photometric selections alone, and have been shown to be underrepresented in spectroscopic samples of high-redshift AGN (e.g. \citealt{hainline_investigation_2025}, \citealt{scholtz_jades_2025}). As such, samples of compact REGs may provide a way to unlock samples of high-redshift AGN which are otherwise challenging to robustly select.

\begin{figure*}
    \centering
    \includegraphics[width=0.95\linewidth]{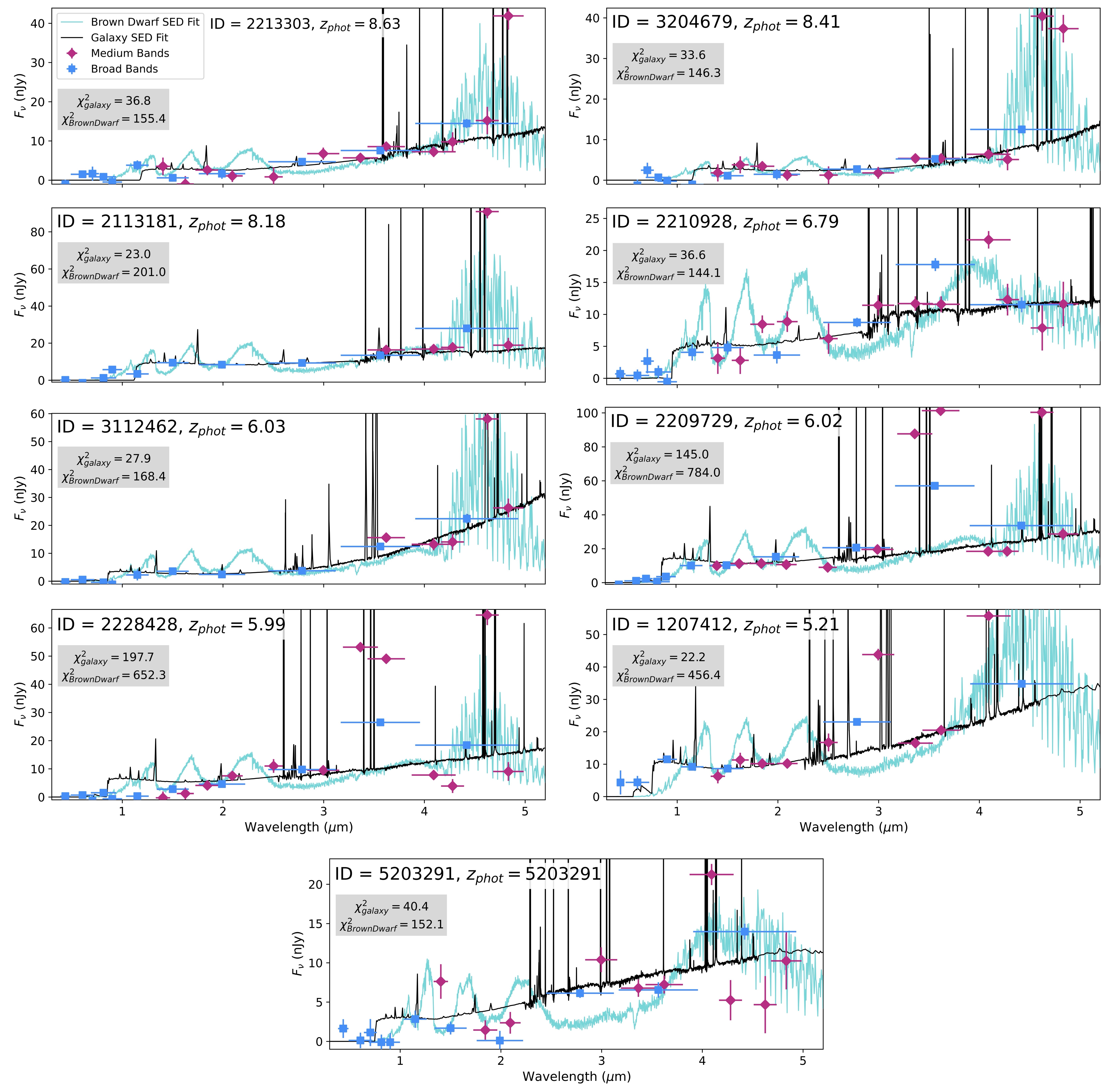}
    \caption{\texttt{EAZY-py} SED fits and photometry for the nine compact REGs. The black curve shows the best-fit galaxy SED based on the \cite{larson_spectral_2023} templates with the \cite{asada_improving_2025} prescription for IGM/CGM attenuation. The turquoise curve shows the best-fit brown dwarf SED based on the Sonora brown dwarf templates (\citealt{marley_sonora_2021}), which include SEDs for extremely cold brown dwarfs ($T_{\text{eff}} \geq 200$K). Each panel shows the $\chi^2$ values for the best-fit galaxy ($\chi_{galaxy}^2$) and brown dwarf ($\chi_{BrownDwarf}^2$) SEDs. All sources are better fit by galaxy templates than brown dwarf templates as required by our selection criteria (see \S \ref{subsec:line_sample}).  }
    \label{fig:compact_REG_SEDs}
\end{figure*}

\section{Compact REGs: Obscured Star Formation, AGN, or A Missing Population of LRDs?} \label{sec:discussion}

In summary, we select a sample of 26 REGs over $5.01 < z < 8.60$. We classify the REGs into three categories based on their colors and morphologies, among which are eight classic LRDs, nine extended REGs, and nine compact REGs. The subsets of REG LRDs and extended REGs have straightforward interpretations. The REG LRDs constitute a relatively normal subset of LRDs (\S\ref{subsec:LRDs}) which have high EW \Ha\ or \OIIIHb\ emission lines. The extended REGs are consistent with DSFGs undergoing merger-induced star-formation (\S\ref{subsec:extended_REGs}). However, as discussed in \S \ref{subsec:compact_REGs_DSFGs} and \S\ref{subsec:AGN}, the interpretation of the compact REGs is not as straightforward. The compact REGs do not qualify as classical LRDs based on the selections of \citealt{kokorev_census_2024} and \citealt{kocevski_rise_2025}, and instead have properties consistent with both AGN and DSFGs. However, neither interpretation is fully consistent with the compact REGs. In this section, we further discuss the interpretation of the compact REGs.

\subsection{Dusty Extreme ELGs or Moderate DSFGs?} \label{subsec:DSFG_discussion}

As discussed in \S \ref{subsec:compact_REGs_DSFGs}, the compact REGs have several characteristics consistent with DSFGs, including physical properties, compact morphologies, and high EW emission lines. When considered individually, the morphologies, red continuum fluxes, and EWs of the compact REGs are consistent with a DSFG interpretation. However, when taken together, the three are difficult to explain in the context of DSFGs. 

To illustrate this, we once again highlight the two key differences between the compact and extended REGs: the compact REGs have significantly lower stellar masses and higher EW emission lines than the extended REGs. The high EWs of the compact REGs make them Extreme ELGs (EELGs), and several of their properties are similar to those of EELGs in the line selected sample. EELGs are often compact, have low stellar masses, high SFRs (Figure \ref{fig:properties}), and EW(\Ha) and EW(\OIIIHb) $\gtrsim 1000$\AA\ (Figure \ref{fig:EW_z}), all of which are observed in the compact REGs. Additionally, EELGs generally have very strong [OIII]$\lambda\lambda4959, 5007$ emission lines, resulting in higher EW(\OIIIHb) $\gtrsim 1000$\AA, and often EW(\OIIIHb)\ $>$ EW(\Ha) (e.g. \citealt{withers_spectroscopy_2023}, \citealt{llerena_extreme_2025}). Such high EW(\OIIIHb) is also observed in seven of the compact REGs.  

However, in contrast to the REGs, EELGs generally have blue continuum fluxes and little dust (\citealt{roberts-borsani_between_2024}, \citealt{llerena_physical_2024}). This would make the compact REGs a unique type of red/dusty EELG. The similarities between the compact REGs and EELGs complicates a DSFG interpretation as it is unclear how the compact REGs are able to maintain such strong emission lines in the presence of $A_V \sim 1 - 2.5$ mag of dust, all contained within $R_{eff}\lesssim0.75 - 1$ kpc. In a DSFG interpretation, this could be explained by complex dust geometries or unique viewing angles of the compact REGs, both of which have been shown to affect how high-redshift reddened/dust obscured galaxies are observed (\citealt{cochrane_disappearing_2024}, \citealt{nelson_jwst_2023}).

As such, if the compact REGs are DSFGs, they would represent a type of lower-mass DSFG, which would qualify as a dusty EELG. Additionally, the faint magnitudes of the compact REGs would place them among the faintest DSFGs. Existing samples of faint DSFGs have been shown to have important contributions to the cosmic SFRD and play an important role in galaxy evolution (\citealt{martis_canucstechnicolor_2025}, COS-z8M1 in \citealt{akins_two_2023}, \citealt{bing_almost_2025}, \citealt{sun_identification_2025}). The compact REGs have several similarities with faint DSFGs, including compactness, $A_V$, and even have similar volume densities ($n \sim 10^{-6} - 10^{-5} \text{Mpc}^{-1}$). However, we once again highlight that the compact REGs have lower stellar masses than existing samples of optically faint DSFGs (\citealt{rodighiero_optically_2024}). This would point to the compact REGs being a distinct type of DSFG than what is commonly reported in the literature, potentially a more moderate type of DSFG, or which has unique dust geometries or viewing angles that allow for the escape of strong emission lines. 

In this context, it is important to highlight the efficiency of the REG selection criteria. The identification of high-redshift, faint DSFGs often relies on MIRI or ALMA observations, which are more limited than NIRCam imaging. Using only imaging from HST and NIRCam, we have potentially identified nine faint DSFGs in less than $77.6 \text{ arcmin}^2$, including three at $z > 7$ where sample sizes are even more limited. Such a selection is challenging without the medium-bands, whose increased spectral resolution enables robust identification of sources with strong emission lines. In fact, all of our compact REGs are robustly identified in the line-selected sample, while only one (CANUCS-2209729) has sufficient continuum \snr\ to be included in the \pz-selected sample. Additionally, only three have sufficient \snr\ in $F444W$ to be selected as EROs based on common selection criteria. The sample of compact REGs is biased towards objects with strong emission lines and is undoubtedly incomplete with respect to galaxies with weak emission lines (such as objects with low SFR). However, this sample provides an efficient way to probe a galaxy population which is otherwise difficult to identify.

\subsection{A Missing Population of LRDs?} \label{subsec:missing_LRDs?}

Perhaps a more natural interpretation for the compact REGs is AGN. An AGN contribution would easily explain their compactness, red continuum colors, and strong emission lines, without requiring potentially complex or rare dust geometries. As discussed in \S \ref{subsec:AGN}, there is evidence for a broad-line component in \Ha\ for the one REG with prism spectroscopy (CANUCS-1207412). This indicates the presence of AGN in at least one of the compact REGs, and suggests AGN is a valid interpretation for the compact REGs. Additionally, we cannot rule out the possibility that the compact REGs have contributions from both star-formation and AGN, which is similar to one interpretation for LRDs (e.g. \citealt{naidu_black_2025}, \citealt{sun_little_2026}, \citealt{merida_testing_2026}).

There are several similarities between the compact REGs and LRDs including compactness, $F150W - F444W$ colors (Figure \ref{fig:ERO_Selection}), $UVJ$ colors (Figure \ref{fig:UVJ}), and a high Balmer decrement. The NIRSpec observations of CANUCS-1207412 yields a Balmer decrement of $\sim 8$, which is similar to those of LRDs reported in the literature (e.g. \citealt{nikopoulos_evidence_2025}). Given the similarities between compact REGs and LRDs, it is worth discussing why the compact REGs are not selected as LRDs. As with \S\ref{subsec:LRDs}, we use the selection of \citealt{kokorev_census_2024} and \citealt{kocevski_rise_2025} for this discussion, and we summarize the results in Table \ref{tab:LRD_criteria}.

\begin{table*}[]
    \centering
    \begin{tabular}{c|c|c|c|c|c|c|c|c|c}
    ID&Type&$z$&EW(\Ha)&EW(\OIIIHb)& $F444W_{mag}$& $A_V$& Stellar Mass                  & SFR                & MWA  \\
      &    &   &  \AA\ & \AA\      &              &  mag &       $\log(M_\star/M_\odot)$ & $\log(M_\odot/yr)$ & Myr \\
    \hline\hline
    2226806	&	LRD	&	5.2	&	875	&	760	&	26.5	&	-	&	-	&	-	&	-	\\
    1200422	&	LRD	&	5.2	&	1019	&	543	&	25.3	&	-	&	-	&	-	&	-	\\
    3108879	&	LRD	&	5.44	&	503	&	-	&	25.6	&	-	&	-	&	-	&	-	\\
    5212979	&	LRD	&	5.61	&	763	&	301	&	25.7	&	-	&	-	&	-	&	-	\\
    5105197	&	LRD	&	5.63	&	564	&	-	&	26.1	&	-	&	-	&	-	&	-	\\
    5205784	&	LRD	&	6.01	&	2237	&	544	&	25.3	&	-	&	-	&	-	&	-	\\
    2115891	&	LRD	&	6.17	&	661	&	559	&	27.7	&	-	&	-	&	-	&	-	\\
    2228326	&	LRD	&	6.34	&	1038	&	632	&	27	&	-	&	-	&	-	&	-	\\
    \hline 
    4222045	&	Extended	&	5.01	&	304	&	450	&	28	&	1.05	&	9.21	&	0.85	&	173.37	\\
    4200621	&	Extended	&	5.06	&	753	&	527	&	25.6	&	1.56	&	9.57	&	2.06	&	21.72	\\
    5212433	&	Extended	&	5.23	&	262	&	203	&	24.7	&	1.04	&	10.3	&	1.96	&	114.13	\\
    5212435	&	Extended	&	5.23	&	489	&	229	&	27.8	&	0.97	&	9.36	&	1.11	&	124.26	\\
    2201093	&	Extended	&	5.99	&	513	&	150	&	26.4	&	1.44	&	10.13	&	2.04	&	87.73	\\
    2201097	&	Extended	&	6.02	&	380	&	352	&	26.4	&	1.58	&	9.26	&	2.19	&	6.34	\\
    2201227	&	Extended	&	6.03	&	641	&	641	&	25	&	1.58	&	9.65	&	2.65	&	3.3	\\
    2217931	&	Extended	&	6.34	&	700	&	186	&	27.4	&	2.87	&	9.72	&	1.81	&	56.4	\\
    2200880	&	Extended	&	6.5	&	341	&	385	&	26.6	&	0.91	&	9.56	&	1.5	&	77.11	\\
    \hline 
    5203291	&	Compact	&	5.15	&	1173	&	434	&	28.5	&	1.88	&	8.05	&	0.99	&	3.82	\\

    1207412	&	Compact	&	5.21	&	1327	&	843	&	27.5	&	1.11	&	8.03	&	0.87	&	8.09	\\
    2228428	&	Compact	&	5.99	&	1492	&	2278	&	28.2	&	1.59	&	8.1	&	1.11	&	1.7	\\
    2209729	&	Compact	&	6.02	&	1006	&	2116	&	27.6	&	1.19	&	8.16	&	1.17	&	1.84	\\
    3112462*	&	Compact	&	6.03	&	572	&	272	&	28	&	2.64	&	8.11	&	0.4	&	32.16	\\
    2210928*	&	Compact	&	6.79	&	-	&	1398	&	28.7	&	1.02	&	8.25	&	0.67	&	24.25	\\
    2113181	&	Compact	&	8.18	&	-	&	1457	&	27.8	&	0.97	&	7.96	&	0.91	&	5.18	\\
    3204679	&	Compact	&	8.41	&	-	&	1080	&	28.7	&	2.22	&	8.16	&	1.1	&	4.4	\\
    2213303	&	Compact	&	8.63	&	-	&	983	&	28.5	&	2.05	&	8.36	&	1.14	&	8.39	\\

    \end{tabular}
    \caption{Physical properties of the 17 REGs, classified into LRDs, extended, and compact sources as described in \S \ref{subsec:LRDs} and \S \ref{subsec:extended_REGs}. Photometric redshifts are measured using \texttt{EAZY} (\S \ref{subsec:line_sample}), equivalent-widths are measured directly from the medium-band photometry (\S \ref{subsec:measuring_properties}), the F444W magnitudes are measured from the $0.3''$ aperture, and the dust attenuation, stellar masses, star formation rates (SFR) and mass-weighted ages (MWA) measured using \texttt{BAGPIPES} (\S \ref{subsec:measuring_properties}). We omit the physical properties of the LRDs as there is no universally accepted way to measure their physical properties using \texttt{BAGPIPES}. Objects marked with an asterisk are experiencing lensing magnification, with $\mu \sim 1 - 2$, we report the magnification corrected physical properties. }
    \label{tab:properties}
\end{table*}

Most of the compact REGs fail two or three of the LRD selection criteria. Roughly half of the compact REGs (5/9) qualify as LRDs in principle as they have sufficiently blue UV colors and red optical colors. However, they fail the selection due to quality cuts, including cuts on UV or optical \snr. The LRD selections utilize several \snr\ cuts on broad-band filters to ensure sources are well detected in the rest-frame UV and optical, as well as an additional cut on \snr($F444W$). Half of the REGs (5/9) fail at least one of these \snr\ cuts, most commonly cuts on the rest-frame UV \snr. 

Additionally, four of the compact REGs are excluded from the sample of LRDs owing to their high EW emission lines. This cut is meant to remove objects with intrinsically blue continuum fluxes that have red broad-band colors only due to strong emission lines. However, this cut can also exclude objects which have intrinsically red continuum fluxes \textit{and} strong emission lines, as is the case for four of our compact REGs. This is most obvious for CANUCS-2209729 and CANUCS-2228428 (see Figure \ref{fig:compact_REG_SEDs}). Both objects have blue $F356W - F444W$ colours due to the presence of very strong \OIIIHb\ emission lines boosting the $F356W$ flux. This excludes them from our sample of LRDs, which require red $F356W - F444W$ colors. However, with the increased spectral resolution of the medium-bands we are able to show that these objects have intrinsically red continuum fluxes, similar to those of LRDs. Additionally, four compact REGs fail at least one of the UV or optical color cuts, having UV continuum fluxes too red and/or optical colors too blue to qualify as LRDs.

Finally, we highlight that three of the compact REGs fail a color cut meant to exclude brown dwarf contaminants. These cuts exclude objects which have very blue $\sim 1 - 2\mu$m colors (e.g. $F115W - F200W > -0.5$ color in \citealt{kokorev_census_2024}), as brown dwarfs with similarly blue UV colors have been shown to contaminate samples of LRDs (e.g. \citealt{burgasser_uncover_2024}, \citealt{greene_uncover_2024}). While many of the compact REGs have very blue $\sim 1 - 2\mu$m colors similar to brown dwarfs, the increased spectral resolution of the medium-bands allows us to confidently rule out brown dwarf solutions for all the compact REGs. As discussed in \S \ref{subsec:line_sample}, sources must be better fit by galaxy templates than brown dwarf templates to be included in the line-selected sample. Visually inspecting the photometry of the compact REGs, it is clear why they are not well fit by brown dwarf templates. The compact REGs have strong flux excesses in the medium bands, with fluxes $\sim 3 - 6 \times$ higher in the medium-band containing an emission line as opposed to neighbouring medium-bands. While molecular absorption features in brown dwarf atmospheres can cause ``bumps and wiggles'' in the medium-bands, the sharp and extreme flux excesses observed in the compact REGs are much more likely to originate from a strong emission line as opposed to a brown dwarf spectral features. To further highlight this point, we show the galaxy and brown-dwarf SED fits for all nine compact REGs in Figure \ref{fig:compact_REG_SEDs}, and show the $\chi^2$ values for the galaxy and brow dwarf SED fits for each object in Table \ref{tab:LRD_criteria}.

\cite{hviding_rubies_2025} compared a sample LRDs selected though common photometric cuts (\citealt{kokorev_census_2024}, \citealt{kocevski_rise_2025}, and \citealt{barro_extremely_2024}) to LRDs selected through NIRSpec prism observations. They find that photometrically selected samples of LRDs suffer from high levels of incompleteness, which is often driven by broad-band \snr\ cuts or color cuts which are too stringent and exclude LRDs with flatter UV and/or optical continuum fluxes. Many of the compact REGs are excluded as LRDs for similar reasons. Consequently, it is not unreasonable to suggest that our compact REGs are LRDs which are too faint or have broad-band colors which are incompatible with classical LRD selections. 

Additionally, several authors have utilized modified versions of LRD selections to generate broader samples of LRDs and LRD-like objects which fail the classical selection criteria, including objects with very strong emission lines (e.g. \citealt{gentile_not-so-little_2024}, \citealt{akins_cosmos-web_2025}, \citealt{caputi_pseudo_2026}, \citealt{merida_rise_2026}). This has revealed several LRD candidates which may be in transitional evolutionary stages, providing important constraints on the evolution of LRDs.

Given the similarities between the spectral properties of the compact REGs and LRDs, we conclude that the compact REGs are likely a population of LRDs which are ``missing'' from existing datasets for several reasons. In this case, the compact REGs would enable the exploration of several new regions in parameter space. This includes a population of LRDs with unusually high EWs (particularly \OIIIHb\ EWs), the faint end of the LRD population, LRDs which have flatter UV or optical continuum slopes than what is generally required by LRD selections, and LRDs which have very steep UV slopes.

There are several possibilities for the origins of these spectral properties of the compact REGs. One interpretation for LRDs is a two component solution where the optical emission is driven by a Black Hole star (BH*), while the UV emission is driven by a blue star-forming galaxy which hosts the BH* (e.g. \citealt{naidu_black_2025}). The compact REGs could correspond to less extreme types of LRDs which have fainter luminosities than classical LRDs or fall in transitional phases. The different UV and optical slopes of the compact REGs could also be driven by different levels of contribution from the host star forming galaxy or central AGN (e.g. \citealt{hainline_investigation_2025}, \citealt{billand_investigating_2026}). While our compact REGs are far too blue to be BH*'s (left panel of Figure \ref{fig:AGN}), they can provide useful constraints on the evolution of LRDs, including the contribution of the host star-forming galaxy to an LRD. 

Once again, we underscore how critical our sample of medium-band selected ELGs is to the identification of such objects. The increased spectral resolution of the medium-bands is able to easily rule out brown dwarf solutions, without the need for additional color cuts that may exclude unique types of LRDs. Additionally, the medium-bands are able to provide clean measurements of rest-frame optical continuum fluxes, allowing us to identify red objects without needing to account for contamination from strong emission lines, as broad-band selections require. Finally, selecting objects on their emission lines enables the robust selection of very faint objects. Several of the compact REGs (particularly those at $z > 7$) are faint in both the rest-frame UV and optical, which makes them extremely difficult to identify using classical LRD selection criteria. Modified versions of LRD selection criteria making use of the medium-bands may aid in the identification of similar objects in the future. 

\begin{table*}[]
    \centering
    \begin{tabular}{c|c|c|c|c|c|c|c|c|c}
        ID & $\chi_{galaxy}^2$ & $ \chi_{BD}^2$ & UV color & Optical Color & BD Flag & EM Line Flag & UV \snr & Optical \snr & $F444W $ \snr  \\
        
        \hline 
        \hline 
        2213303	&	36.76	&	155.38	&	PASS	        &	PASS	        &	\textbf{FAIL}	&	PASS	        &	\textbf{FAIL}	&	\textbf{FAIL}	&	PASS	\\
        3204679	&	33.65	&	146.34	&	PASS	        &	PASS	        &	\textbf{FAIL}	&	PASS	        &	\textbf{FAIL}	&	PASS	&	PASS	\\
        2113181	&	23.04	&	200.97	&	PASS	        &	\textbf{FAIL}	&	PASS	        &	\textbf{FAIL}	&	PASS	        &	PASS	&	PASS	\\
        2210928	&	36.63	&	144.06	&	PASS	        &	PASS	        &	PASS	        &	PASS	        &	\textbf{FAIL}	&	PASS	&	PASS	\\
        3112462	&	27.87	&	168.42	&	PASS	        &	PASS	        &	PASS	        &	PASS	        &	\textbf{FAIL}	&	PASS	&	PASS	\\
        2209729	&	145.02	&	784.03	&	PASS	        &	\textbf{FAIL}	&	PASS	        &	\textbf{FAIL}	&	PASS	        &	PASS	&	PASS	\\
        2228428	&	197.66	&	652.3	&	\textbf{FAIL} 	&	\textbf{FAIL}	&	PASS	        &	\textbf{FAIL}	&	PASS	        &	PASS	&	PASS	\\
        1207412	&	22.23	&	456.43	&	PASS	        &	\textbf{FAIL}	&	PASS	        &	\textbf{FAIL}	&	PASS	        &	PASS	&	PASS	\\
        5203291	&	40.38	&	152.07	&	PASS	        &	PASS	        &	\textbf{FAIL}	&	PASS	        &	\textbf{FAIL}	&	PASS	&	PASS	\\

    \end{tabular}
    \caption{Summary of the compact REGs and how they compare to commonly used LRD selections. We show the $\chi^2$ values from the \texttt{EAZY-py} SED fits using galaxy templates ($\chi_{galaxy}^2$) and brown dwarf templates ($ \chi_{BD}^2$, see \S \ref{subsec:line_sample} for details on the fitting). The reaming columns show which LRD cuts the compact REGs pass/fail. This includes cuts on the UV and optical colors, the \snr\ of the bands used to measure the UV and optical colors, $\sim1 - 2 \mu$m colors meant to remove brown dwarf contamination and colors meant to exclude sources with strong emission lines. The compact REGs fail LRD selections for a variety of reasons, including several objects which have low \snr, colors incompatible with LRDs, very strong emission lines, and rest-frame UV colors which are too similar to brown dwarfs.  }
    \label{tab:LRD_criteria}
\end{table*}

\section{Summary \& Conclusion}

We use the NIRCam medium-band color selections to identify a sample of Emission Line Galaxies (ELGs) over $4.89 < z < 8.9$. These galaxies exhibit a statistically significant correlation between continuum color and emission line strength, where bluer galaxies have strong \Ha\ and \OIIIHb\ emission lines. We fit this relation with a straight line, and identify 26 Red Emission line Galaxies (REGs), which have continuum colors $\geq 2\sigma$ redder than the typical continuum color based on the fit. Using the selection criteria from \cite{kokorev_census_2024} and \cite{kocevski_rise_2025}, we classify eight REGs as LRDs (the REG LRDs, \S \ref{subsec:LRDs}), which appear to constitute a relatively normal type of LRD. We then classify the remaining REGs into extended and compact sources based on their $F444W$ morphologies. The extended REGs (nine in total) are resolved in $F444W$ and have $1 \lesssim R_{eff} \lesssim 3$ kpc while the compact REGs (nine in total) are unresolved in $F444W$. The extended REGs appear to be DSFGs undergoing merger-induced star-formation (\S\ref{subsec:extended_REGs}). However, the nature of the compact REGs is more puzzling, as they have properties consistent with DSFGs, AGN, and classical LRDs. We discuss three interpretations for the compact REGs:

\begin{itemize}
    \item \textbf{DSFG interpretation, \S\ref{subsec:compact_REGs_DSFGs}, \S\ref{subsec:DSFG_discussion}:} if the compact REGs are DSFGs, their faint NIRCam magnitudes suggest they are faint DSFGs which are challenging to identify using common DSFG selection criteria. In this interpretation, the compact REGs would represent a unique type of DSFG which have lower stellar masses and higher EW emission lines than typical samples of DSFGs. Several of their properties are consistent with EELGs, making the compact REGs a type of dusty EELG. However, the combination of the compact morphologies, strong emission lines, and red continuum fluxes are challenging to explain in the context of DSFGs, and likely rely on the presence of complex or rare dust geometries. 

    \item \textbf{AGN interpretation, \S\ref{subsec:AGN}:} the presence of AGN in the compact REGs would provide a natural explanation for their compact morphologies, strong emission lines, and red continuum fluxes. One compact REG (CANUCS-1207412) has available NIRSpec prism spectroscopy (Figure \ref{fig:nirspec}) and shows evidence of broad \Ha, supporting the AGN interpretation. Comparing the compact REGs to a sample of AGN with broad \Ha\ (Figure \ref{fig:AGN}) reveals the compact REGs have generally fainter $M_V$, bluer $\beta_{tot}$, and higher EW(\Ha) and EW(\OIIIHb) than existing samples of AGN. Thus, if the compact REGs are AGN, they could represent low-luminosity AGN which have been shown to be underrepresented in exiting spectroscopic samples of high-redshift AGN (e.g. \citealt{hainline_investigation_2025}). 

    \item \textbf{LRD interpretation, \S\ref{subsec:missing_LRDs?}:} the compact REGs share several similarities with LRDs, including compact morphologies, high EW emission lines, broad-band colors, and a large Balmer decrement in the one compact REG with available spectroscopy. We investigate why the compact REGs are not classified as LRDs (summarized in Table \ref{tab:LRD_criteria}), and find that the compact REGs fail a variety of LRD selection criteria. These reasons include objects which are too faint, have continuum fluxes too flat to meet classical LRD selection criteria (similar to the results of \citealt{hviding_rubies_2025}), and several objects that have very strong emission lines which generate blue rest-frame optical colors in the broad-bands. Additionally, many of our compact REGs have $\sim 1 - 2\mu$m colors which are similar to brown dwarfs. This rules them out as LRDs based on many common selection criteria, however, brown dwarf solutions are ruled out for the compact REGs thanks to the increased spectral resolution of the of the medium-bands. If the compact REGs are LRDs, they may enable the exploration of LRDs in different evolutionary stages than classical LRDs, or less extreme versions of LRDs which are fainter or have less extreme colors than classical LRDs. 
    
\end{itemize}

Based on the similarities between the compact REGs and LRDs, we conclude that the compact REGs are most likely LRDs which fail classical LRD selection criteria. However, in all three interpretations, the compact REGs would represent unique types of galaxy populations which are largely missing from existing datasets. Further constraints (such as NIRSpec spectroscopy or MIRI photometry) on these objects could provide a wealth of new information regarding the evolution of LRDs, as well as potentially obscured star formation or AGN activity at high-redshift. 

Finally, we highlight how critical the medium-bands are to the identification of the compact REGs. Objects such as these remain difficult to robustly identify owing to their faint magnitudes ($F444W > 27$ mag and often $F444W > 28$ mag, Table \ref{tab:properties}), meaning similar objects are largely missing from both spectroscopic and photometric galaxy samples. However, they can be easily selected when targeting strong emission lines in the medium-bands.


\begin{acknowledgments}
This research was supported by grants 18JWST-GTO1,  23JWGO2A13, 23JWGO2B15 and  24JWGO2A04 from the Canadian Space Agency (CSA) and funding from the Natural Sciences and Engineering Research Council of Canada (NSERC). AM acknowledges support from the Yavin Family Fund.  The CANUCS observations used in this paper are available on the Mikulski Archive for Space Telescopes (MAST) at the Space Telescope Science Institute, which can be accessed via doi:10.17909/ph4n-6n76.  This research used the Canadian Advanced Network For Astronomy Research (CANFAR) operated in partnership by the Canadian Astronomy Data Centre and The Digital Research Alliance of Canada with support from the National Research Council of Canada, the Canadian Space Agency, CANARIE, and the Canadian Foundation for Innovation. DM acknowledges generous support from the Leonard and Jane Holmes Bernstein Professorship in Evolutionary Science. Support for programs JWST-GO-03362 and JWST-GO-05890, provided through a grant from the STScI under NASA contract NAS5-03127, is acknowledged.
\end{acknowledgments}


%






\bibliographystyle{aasjournalv7}
\bibliography{REG_v3}{}



\end{document}